\documentclass[acmsmall,screen]{acmart}

\usepackage{graphicx}
\usepackage{listings}

\newcommand{\ignore}[1]{}
\newcommand{\desc}[1]{}

\newtheorem{assumption}{Assumption}
\newtheorem{claim}{Claim}

\makeatletter\if@ACM@journal\makeatother
\setcopyright{rightsretained} 
\acmJournal{PACMPL}
\acmYear{2017} \acmVolume{1} \acmNumber{1} \acmArticle{67} \acmMonth{1} \acmPrice{}\acmDOI{10.1145/3133891}
\fi

\begin{document}

    \title{Efficient Logging in Non-Volatile Memory by Exploiting Coherency Protocols}         


    \author{Nachshon Cohen}
    \orcid{0000-0001-8302-2739}             
    \affiliation{
        \institution{EPFL}            
        \city{Lausanne}
\country{Switzerland}
    }
    \email{nachshonc@gmail.com}          

    \author{Michal Friedman}
    \affiliation{
        \institution{Technion}            
    }
    \email{michal.fman@gmail.com}          

    \author{James R. Larus}
    \affiliation{
        \institution{EPFL}            
        \city{Lausanne}
        \country{Switzerland}
    }
    \email{james.larus@epfl.ch}          

    \thanks{The paper will appear at Proceedings of the 2017 ACM SIGPLAN International Conference on Object-Oriented Programming, Systems, Languages, and Applications (OOPSLA'17). ACM,  Vancouver, Canada.}                

    \begin{abstract}
Non-volatile memory (NVM) technologies such as PCM, ReRAM and STT-RAM allow processors to directly write values to persistent storage at speeds that are significantly faster than previous durable media such as hard drives or SSDs.
Many applications of NVM are constructed on a logging subsystem, which enables operations to appear to execute atomically and facilitates recovery from failures.
Writes to NVM, however, pass through a processor's memory system, which can delay and reorder them and can impair the correctness and cost of logging algorithms.

Reordering arises because of out-of-order execution in a CPU and the inter-processor cache coherence protocol.
By carefully considering the properties of these reorderings, this paper develops a logging protocol that requires only one round trip to non-volatile memory while avoiding expensive computations.
We show how to extend the logging protocol to building a persistent set (hash map) that also requires only a single round trip to non-volatile memory for insertion, updating, or deletion.
    \end{abstract}




    \maketitle

\def \set {single-trip persistent set}
\def \Set {Single-trip persistent set}
\def \sset {{\sc STPS}}

\section{Introduction}

\desc{Introduce NVM and its "popularity"}
New memory technologies are changing the computer systems landscape.
Motivated by the power and volatility limitations of Dynamic Random Access Memory (DRAM), new, non-volatile memory (NVM) technologies -- such as ReRAM \cite{Akinaga2010,Wong2012}, PCM \cite{Qureshi2009,Lee2009}, and STT-RAM \cite{Hosomi} -- are likely to become widely deployed in server and commodity computers in the near future.
Memories built from these technologies can be directly accessible at the byte or word granularity and are also non-volatile.
Thus, by putting these new memories on the CPU's memory bus, the CPU can directly read and write non-volatile memory using load and store instructions.
This advance eliminates the classical dichotomy of slow, non-volatile disks and fast, volatile memory, potentially expanding use of durability mechanisms significantly.

Taking advantage of non-volatility is not as simple as just keeping data in NVM.
Without programming support, it is challenging to write correct, efficient code that permits recovery after a power failure since the restart mechanism must find a consistent state in durable storage, which may have last written at an arbitrary point in a program's execution.
This problem is well-known in the database community, and a significant portion of a DB system is devoted to ensuring durability in the presence of failures.
NVM is different, however, because writes are fine-grain and low-cost and are initiated by store instructions.
A further complication is that a processor's memory system reorders writes to NVM, making it difficult to ensure that program state, even when consistent in memory, is recorded consistently to durable storage.
In the interest of high performance, processors employ caches and write buffers and store values to memory at unpredictable times.
As a consequence, stored values may not reach NVM in the same order in which a program executes them, which complicates capturing a consistent snapshot in durable storage.

To simplify software development, most programming constructs for NVM provide all-or-nothing semantics: either all modifications in a block survive a crash or none of them do.
To implement these all-or-nothing blocks, most systems use either undo or redo logging (cf Section~\ref{sec-related}).
Logging is a fundamental operation to use NVM effectively.
With undo logging, the log holds the old value from each modified location, which suffices to restore the system to the state before a partially executed block.
In redo logging, all modifications are stored directly in the log.
Once a block is complete, the log contains sufficient information to fully replay the block.
In both cases, every write executed in an all-or-nothing block modifies the log, so its write performance is fundamental to system speed.

The most important operation supported by a log is {\em appending} a data item (an {\em entry}) to the tail of the log.
It is also possible to read from the log and to trim part of the log, but these operations typically occur outside of the critical path of execution.
Once an entry is appended to the log, it must remain visible after a power failure and system restart (i.e., it is {\em persistent}).
Entries that were only partially recorded must not be recovered after a power failure since their content is corrupt.
The primary challenge in designing a log data structure is distinguishing between fully and partially written entries.

\desc{Describe the standard solution and the overhead of two round trips to memory.
    It also hints the main optimization we are going to use: writing an entry in a single trip to NVM.}
A standard solution is to split an append operation into two steps.
In the first, the entry itself is written and flushed to NVM \cite{Schwalb}.
In the second step, the data is committed, or made visible, by inserting a commit record or by linking the data to a list.
When the second write reaches NVM, the entry is guaranteed to fully reside in NVM.
This approach requires at least two round-trip accesses (each a store, a flush, and a fence) to NVM.
Even though an NVM access is far faster than a disk or SSD, it still crosses the memory bus and is one to two orders of magnitude slower than references to the processor cache\footnote{\citet{Bhandari2016,Chakrabarti2014} reported 200ns latency for the {\sf clflush} x86 instruction.}.
Therefore, it is desirable to reduce cost of durability by making only a single round trip to NVM.

\desc{Describe some ad-hoc solutions that avoid two round trips to memory.
  It shows that this optimization is important.
  Also, it shows that existing solution has non-trivial costs.}
\desc{Describe the key point of this article: using the characteristics of the persistent memory coherency protocol.}
There have been many attempts to reduce the cost of flushes to NVM (Section~\ref{sec-related}).
In this paper, we propose an alternative solution that depends on two properties of modern processors: 1) the cache line granularity of transfers from cache to NVM and 2) the ability to control the order of cache accesses.

The key observation is the last store to a cache line is sent to memory no earlier than previous stores {\em to the same cache line}.
Thus, to distinguish a fully written cache line from a partially written cache line, it suffices to determine if the last write made it to memory.
Using this observation, we propose a log algorithm that always avoids the second round trip to NVM.
The algorithm is easy to deploy, supports different entry sizes, and does not require new hardware support.

\desc{And our solution is not just theoretical: it really works and improves performance on actual system.
  Also, it is easy to employ (add after we have some data).}
We tested the effectiveness of our solution by implementing a log to support transactional memory.
We replaced the logging algorithm in the Atlas \cite{Chakrabarti2014} NVM programming language by ours, which improved performance on micro-benchmarks by up to 38\%.
We also modified TinySTM \cite{Felber2008}, a software transaction memory system, to add persistent transactions on NVM.
The new logging algorithm improved performance by up to 42\%.

We then extended the log algorithm to build a persistent NVM set (hash map).
This persistent set is able to persist new data with a single round trip to NVM, which both improves throughput and reduces the risk of losing data.
Furthermore, it also allows a limited form of transaction, while still requiring only one round trip to NVM.

\section{Memory Coherency Protocol for Non-Volatile Memory}\label{sec-memoryorder}
\desc{Introduce the structure of the chapter}
In this section, we explore the interaction between memory coherency protocols and NVM.
There are two relevant protocols: the standard CPU-cache memory coherency protocol and the cache-NVM protocol.
We discuss these protocols and the result of their composition.

\subsection{Protocol Between CPU and Cache}
The memory coherency protocol among the CPU, cache, and memory has been widely studied in the context of parallel programs, as shared-memory communication is effected by inter-cache transfers.
To ensure that concurrent threads (which possibly run on different processors) observe state modifications in the desired order, modern programming languages, such as C++11, provide explicit memory reference ordering statements.
They enable a programmer to constrain the order in which stores reach the cache subsystem.\footnote{In this paper, we
    are only concerned the cache and its coherency protocol, and not, for example, store buffers.
    We only consider stores that reach the cache and not cache-bypassing stores.}

Specifically, a write with release memory ordering semantics ensures that its value is visible to other threads later (or at the same time) than values from writes that executed previously.
We also require the more expensive operation: release memory fence.
It ensures that values from any write that executed after the fence are visible to other threads later (or at the same time) than values from writes that executed previously\footnote{In
    C++11, these operations are {\sf atomic\_store\_explicit(addr, value, memory\_order\_relaxed)} and {\sf atomic\_thread\_fence(memory\_order\_release)}, respectively.
    The latter imposes more restrictions on a compiler since it affects any write following the fence while the former restricts only a single write.
    Java and C\# guarantee that two volatile writes reach the cache in order of execution.}.
On x86 processors, these are compiler directives that do not incur runtime overhead (beyond reduced opportunity for compiler optimizations).

Memory ordering directives specify the order in which writes become visible to other threads.
They do not specify the order in which the writes reach caches, since a cache is an implementation mechanism, generally invisible in language specifications.
We make an assumption that constraining the order of writes with respect to other threads also constrains the order in which the writes reach the cache.
This assumption is reasonable for existing processors,
since writes only become available to other processors after they are stored in the cache.
\begin{assumption}\label{assumption-multithread}
    If two stores $X$ and $W$ are guaranteed to become visible to concurrent threads in that order, then they are guaranteed to reach the cache subsystem in the same order.
    Hence the memory ordering directives can be used to control the order in which stores reach the cache.
\end{assumption}

\subsection{Protocol Between Cache and NVM}
\desc{Why we care about memory coherency of cache?}
In existing computers, the processor cache is volatile.
The durability of a write is ensured only when its cache line is written or flushed to NVM itself.

\desc{Memory coherency of cache: any reordering is possible.
  Defining  "moving from one memory to another", internal moves (external are discussed below), and "unit of transfer".}
Cache lines are written back to memory in accordance with the cache's policy.
In effect, this means that there is no ordering constraint on writes to NVM.
Modified ({\em dirty}) cache lines can be written to NVM in any order.
However, current systems do not write a partial cache line; every transfer moves an entire cache line to the memory.
We assume that this data transfer is atomic, so that multiple modifications to the same cache line are either fully written or not written at all when the line is flushed to memory after the writes.
This assumption is also used by \citet[NPM Annex B]{SNIA2013} and \citet[see footnote 16]{Chakrabarti2014}.
\begin{assumption}\label{assumption-atomicity}
    A single cache line is transferred from the cache to NVM atomically.
    Thus, if a range of memory words is contained in a single cache line and the data stored in the cache differs from the corresponding values stored in memory, then after a crash, NVM can contain either the memory version or the cached version, but not a mixture of both.
\end{assumption}

It is important to note that multiple writes to the same cache line {\em are not} executed atomically from the perspective of NVM since these writes do not reach the cache atomically.

\newcommand{\clflush}[1]{{\sf clflushopt}(#1)}
\def \mfence {{\sf sfence}}
\def \sfence {{\sf sfence}}
\def \cto {{${\mathcal P}$CSO}}

\desc{Explicit flush to NVM}
Dirty cache lines are flushed to memory either internally, because of cache write backs, or explicitly, because of flush instructions.
Some systems provide operations to flush the entire cache (e.g., x86 {\sf wbinvd} instruction), but this is a very expensive operation and should generally be avoided.
Instead, we rely on the possibility of flushing {\em specific} cache lines.
We further assume that a flush operation executes asynchronously, so that multiple cache line flushes can be pipelined.
To ensure that all outstanding flushes reach memory, a fence operation is necessary.
Throughout this paper we use the x86 terminology: {\sf clflushopt} flushes a cache line asynchronously and {\sf sfence} blocks until all previously executed flushes complete.
Equivalent ARM instructions also exist\footnote{http://infocenter.arm.com/help/index.jsp?topic=/com.arm.doc.den0024a/BABJDBHI.html}.
As we will see, it is beneficial to write the content of a cache line to memory without evicting it from the cache, so it can be reaccessed at low cost.
The x86 {\sf clwb} instruction provides this functionality, but it is not yet available on x86 processors.
The ARM64 cache flush operations already provide this option.

\subsection{Protocol Between CPU and NVM}
The coherency protocol between the program (CPU) and the memory is formed by the composition of these two protocols.
We are mainly concerned with writes, as reads are generally unaffected by the non-volatility of the memory.

We use the happens-before relation of C++ and Java memory models \cite{Manson2005,Boehm2008}\footnote{We do not consider memory\_order\_consume due to subtleties that are outside the scope of this paper.}, denoted by $<_{hb}$.
According to Assumption~\ref{assumption-multithread}, given two write operations $W$ and $X$ with $W<_{hb}X$, then $W$ reaches the cache before (or at the same time) $X$ reaches the cache. 
Given a write operation $W$, we denote its write address by $a(W)$ and the address of the cache line by $c(a(W))$, or directly by $c(W)$. 
Given the address of a cache line $C$,
we denote by \clflush{C} a library call that flushes cache line $C$ asynchronously. 
We denote by \sfence{}() a library call that waits until all asynchronous flushes, which are executed by the caller thread, are completed. 
%

We define {\em persistent ordering}, $X<_pW$ if
$X$ is written to persistent memory no later than $W$.
Then the following holds:
\begin{itemize}
    \item $W<_{hb} \mbox{{\em clflushopt}}(c(W))<_{hb}\mbox{{\em sfence}}()<_{hb}X \Rightarrow W<_pX$ (explicit flush).
    \item $W<_{hb}X \wedge c(W) = c(X) \Rightarrow W<_pX$ (granularity).
\end{itemize}
We denote the resulting persistent memory coherency protocol by ${\mathcal P}$CSO for {\em Persistent Cache Store Order}.

\def \false {{\sc false}}
\def \true {{\sc true}}
\def \invalid {{\sc invalid}}
\def \valid {{\sc valid}}

\section{Log Algorithms}\label{section-algorithm}

The key idea of our log algorithms is to distinguish if the
last word written into a cache line, or even last bit written into a cache line, has made it to NVM or not.
The CPU-cache protocol ensures an ordering of writes within the cache line, and consequently enables us to uniquely distinguish the last write.
Afterwards, the algorithm can flush the line to NVM.
According to the ${\mathcal P}$CSO consistency model,
if the last write made it to NVM, then all preceding writes to the cache line also made it to NVM.

Specifically, we consider two cases.
In the first case, the log algorithm metadata and the log entry can fit in the same cache line.
In this case, the algorithm writes the metadata immediately after writing the actual data.
A cache line in NVM is valid if its metadata contains the sentinel value {\sc valid}.
A log entry is valid if all cache lines on which the entry resides are valid. It is only partially recorded otherwise.
When the log's memory is reused, we can swap the meaning of {\sc valid} and {\sc invalid} swap, to avoid the need to reinitialize the log.

In the other case, where data is consecutive in memory with no space for metadata, we present two variants of the algorithm that allow metadata to be stored separated from data, either by carefully using part of the data as a validity bit or by employing randomization.

\subsection{Algorithm Details}
\def \tworounds {TwoRounds}
\def \atlaslog {AtlasLog}
\def \vb {{CSO-VB}}
\def \random {{CSO-Random}}
\def \fvb {{CSO-FVB}}

\desc{Section layout. Define the problem.}
A log supports two operations that modifies the log: one for appending a log entry to the log tail and another for trimming a set of entries from the log head.
We assume that the performance of appends is more important than trimming, as appends execute on the critical path, while trimming may happen asynchronously.\footnote{The frequency of reading from the log depends on the actual usage.
    In some systems, the log is read only after a power failure and during recovery.
    In other systems, the log is read before entries are trimmed, for example to flush their content to NVM.
    It is also possible that some operation read the log during normal execution, such as for redo logging systems, but this should happen infrequently.}

We further assume that the log is implemented as a circular buffer (or, a set of connected circular buffers), which are reused.
Finally, we assume that a cache line is 64 bytes (512 bits) and a machine word is 8 bytes (64 bits).

A log append operation receives the {\em data} or {\em payload} to append.
The operation creates a {\em log entry} that contains the payload and additional {\em metadata}.
A system may crash (e.g., due to a power failure) at any point during execution.
During recovery, it must be possible to distinguish between the case in which the payload was fully written to NVM from the case in which only part of the payload was stored.
When the append operation returns, it is guaranteed that the data is stored durably and the recovery procedure will recognize this state.
In addition to correctness, the log algorithm should perform well, minimizing the latency for the append operation.

\desc{Multithread support for the log.}
The algorithms below do not contain explicit synchronization.
Sharing a log among threads and serializing writes with a global lock can significantly reduce performance \cite{Avni2016,Wang2014}.
There are better ways to parallelize a log.
One option is to partition the log, which allows threads to write simultaneously to the log without synchronization.
This typically requires a shared variable, accessed with a fetch-and-add instruction, to obtain a new portion of log space.
Another possibility is a per-thread log augmented by additional ordering information (such as a shared sequence number) in the log entry.
In this case, each log is private, but the global order of entries can be reconstructed.
We use the latter option for our logging algorithm in the Atlas and TinySTM systems.

\subsubsection{Validity Bit}
\label{sec:vb}
First, consider the case when it is possible to dedicate some metadata bits in the circular log.
Specifically, we require that every cache line spanned by any log entry contains at least one metadata bit at a known location.

If the size of an payload is small, it is possible to add a metadata byte or word after the data.
For example, if a payload contains 24 bytes of data, then an additional metadata word can be appended after it.
Thus, every fourth word in the log is metadata.
When the log is aligned with cache lines, then every log entry fits in a single cache line and contains one metadata word.

\begin{figure}
    \noindent\makebox[90mm]{
        \includegraphics[height=24mm]{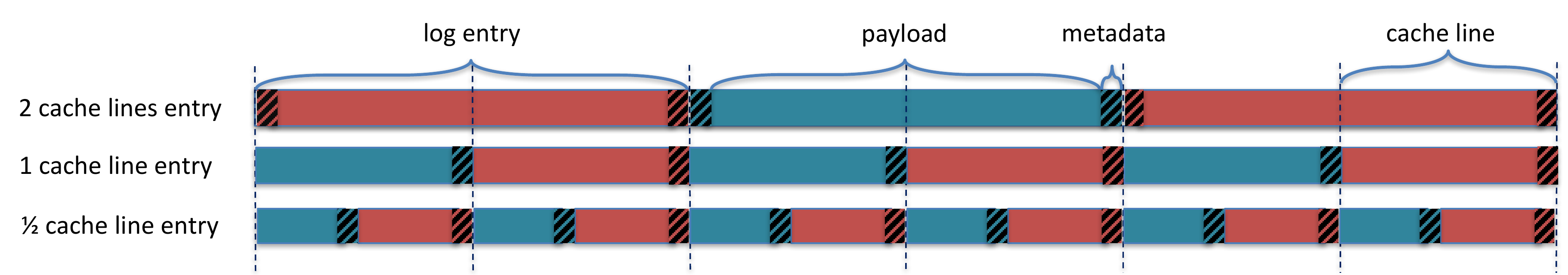}}
    \caption{Different alternatives for different sized log entries.
        Vertical lines represent cache-line boundaries.
        The full area represents payload while the dash area represents validity bits.
        Colors are used to distinguish consecutive entries.}
    \label{figure:validatedWord}
\end{figure}

Generally, it is possible to place metadata bits inside the log if the size of a payload is smaller than 2 cache lines minus two words.\footnote{We assume that the content must be word-aligned so that the application can read the log easily.
    Thus, the payload cannot start at the second byte (or bit) and must start at the second word.
    In the common case, the size of the payload is also word-aligned. Thus, the second validity bit must also reside in a dedicated metadata word. Overall, 2 metadata words are used.
    If the size is not word-aligned, it is possible to use one word and one byte for metadata.
    If the payload is not required to be word-aligned, it is possible to use only two bytes for metadata.}
In this case, it is possible to expand each entry to one or two cache lines and place a metadata word at the beginning of a cache line and another metadata word at the end of the subsequent cache line.
Figure~\ref{figure:validatedWord} illustrates how different object sizes are handled.

In other cases, if the structure of the payload is known, it may be possible to use unused bits in the payload as validity bits.
For example, if the payload contains a value and an address,
and the address is known to be at least 2-byte aligned,
then the address's least significant bit (which is zero) can hold the validity bit.
Although bit manipulations are slightly more complex, this approach has the benefit of minimizing memory footprint.

Formally, the algorithm works only if the condition below is satisfied.
\begin{definition}
    For any log entry $L$ and every cache line $C$ overlapped by $L$, there must be a metadata bit in $C\cap L$.
\end{definition}

The proposed log algorithm is called \vb{}.
The log is initialized to zero and a zero validity bit implies \invalid{}.
When a new entry is added, the data is written (but not flushed).
Then, all validity bits inside the entry are set to \valid{} (initially 1).
Then the entries are flushed.
When the log is reused (i.e., tail reaches index 0 again), the meaning of the validity bit is flipped, so that 0 implies \valid{} and vice versa.

Since the validity bit is written after the payload, according to the {${\mathcal P}$}CSO consistency model, a validity bit containing \valid{} implies that the entire cache line was also written to NVM.
Thus, if all validity bits inside an entry are \valid{}, the entire log entry was written.

The tail is kept in volatile memory and is not considered during recovery.
The head pointer is kept in NVM and must also provide the current polarity of the validity bit.
A log entry is \valid{} if it is in the range [head, LOGSIZE) and its validity bit matches the current validity bit,
or it is in the range [0,head) and its validity bit is opposite from the current validity bit.
The entry pointed to by head is the oldest entry.
Entries in [head, LOGSIZE) are ordered by their distance from head (closer means older).
Entries [0,head) are newer than entries in [head, LOGSIZE) and are ordered by their distance from the beginning of the log (closer means older).

\subsubsection{Preserving Entry Layout via Randomization}
\label{sec:random}
While the \vb{} algorithm is efficient in space and time, it requires interleaving validity bits with log entries.
If a log entry is long, using this approach requires splitting an entry into multiple smaller entries.
This may break or require non-trivial modification to existing code.
For example, if a string is written to the log, it is highly desirable to store it continuously in memory so that string-handling methods operate normally.
\random{} is a variant of the algorithm that alleviates this issue (at a cost), to durably store log entries that do not allow modification of their internal layout.

The key idea is to initialize the log memory with a predefined 64-bit random value.\footnote{Another variant is to initialize memory with an invalid value that will later be overwritten by a valid value when the actual data is stored. For example, the upper 16 bits of addresses on 64-bit x86 processors must be either 0 or 1. Setting these bits to another pattern in a pointer field in a log entry can distinguish a valid address from an initialized value.}
When a payload is logged, its last word must differ from the random value.
If the random value is chosen in a uniform manner, the probability of a collision is $2^{-64}$ per cache line of payload.
On recovery, if the value in the appropriate word of NVM cache line differs from the random value, then the cache line was fully written.

On the other hand, in the rare case when corresponding word of the value matches the random value, we require an additional round-trip to NVM.
In our implementation, we assume the existence of a {\sf sentinel} value that cannot appear in normal execution, but differs from the random value.
After writing the log entry, we append another log entry containing the {\sf sentinel} value.
Once this second entry is valid, it serves as an indicator that the first entry is also valid.

Unfortunately with this algorithm, every append requires two round trips to NVM: first to initialize NVM to the random value and the second to actually write the data.
Fortunately, the critical path to append data contains only a single round trip, as the initialization can be done in advance.
In addition, during initialization, many cache lines can be written between each flush instruction to improve performance.
Still, this is not a great solution as NVM is likely to incur high energy consumption during the additional write, which also counts against possible wear limits of NVM.
Below, we consider another solution that avoids the second write.

\subsubsection{Flexible Validity Bit}
\label{sec:fvb}
\begin{figure}
    \includegraphics[width=67mm]{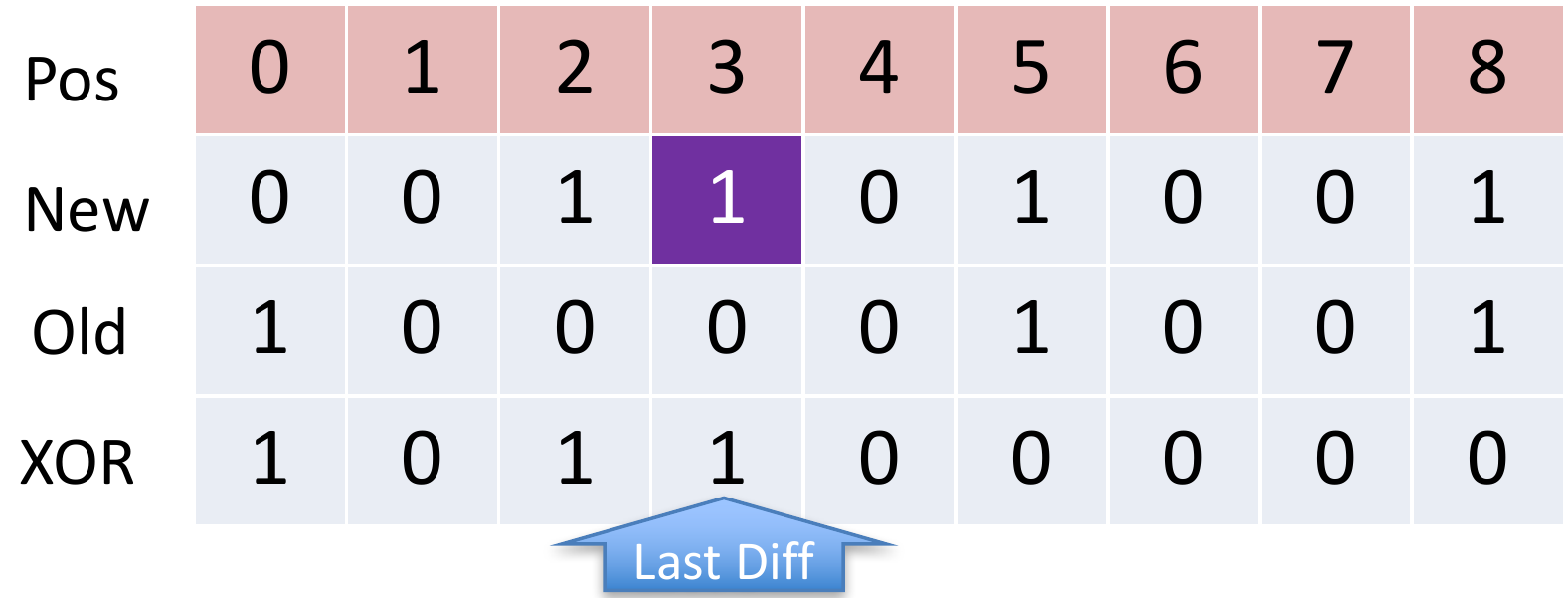}
    \includegraphics[width=67mm]{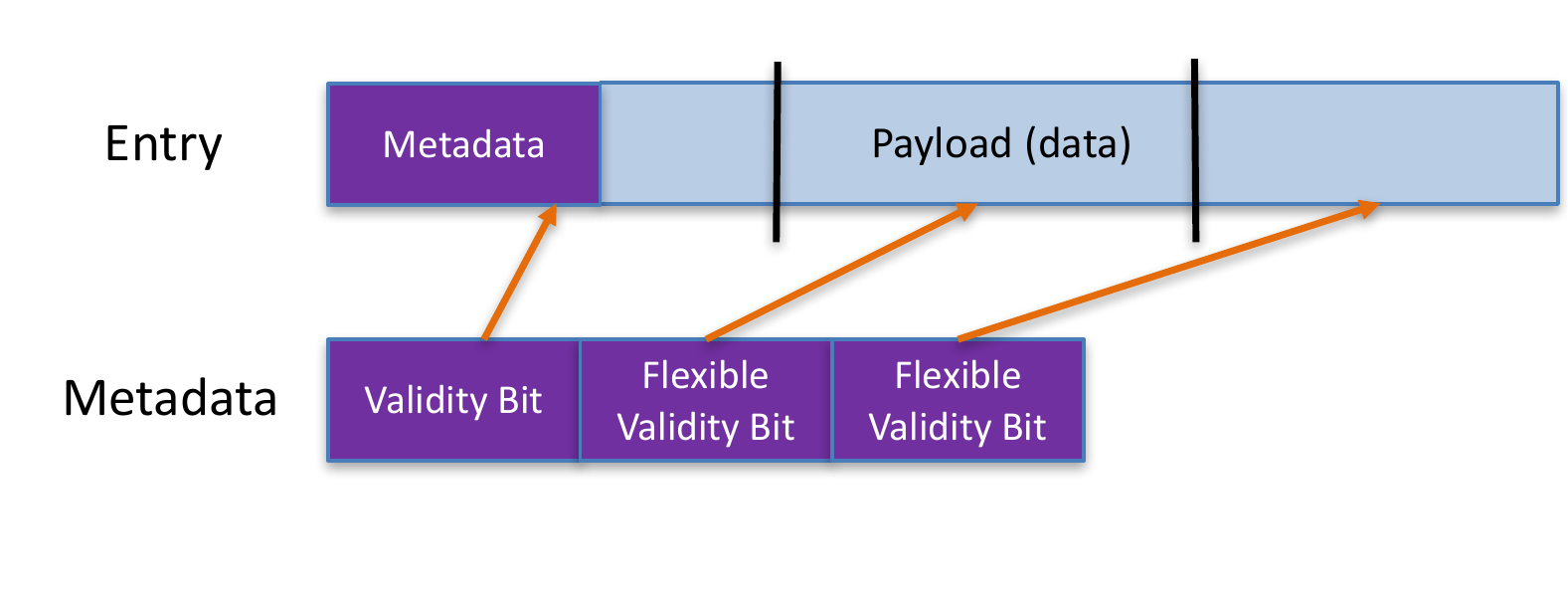}
    \\\noindent (a) \hspace{7cm} (b)
    \caption{
    (a) Single validity bit.
            The position of the last bit that differs between the old and the new content is marked with a blue arrow.
            The bit in the new content at this position is the flexible validity bit; it is marked in purple.
            The position of the flexible validity bit is 3 and its value is 1.
            Part (b) contains a log entry for the flexible validity bit algorithm.
            The payload spans 2 full cache lines and part of the first cache line (up to 7 words).
            There is one metadata word at the beginning of the entry.
            This metadata word contains one validity bit and two flexible validity bits.
            Orange arrows show which validity bits are used to ensure the validity of each cache line in the entry.
    }
    \label{figure:flexible-vb}
\end{figure}

The Flexible Validity Bit (\fvb{}) algorithm is similar to the \vb{} algorithm above: it uses a single validity bit in each cache line to indicates it the cache line was fully written.
However, unlike the algorithm above, it is not possible to ``steal'' a single bit from the actual data.
Thus, both the new data to be written to the log and the old data already in the log are arbitrary bit patterns.
The key idea in the \fvb{} algorithm is to find the last bit that distinguishes between the old and the new data, which then serves as a validity bit.
Finding the position of the flexible validity bit requires a bitwise XOR between the old and new content and then finding the final 1 bit.
The algorithm stores both the position of the validity bit and the value of the bit in a separate metadata field.
Figure~\ref{figure:flexible-vb}(a) illustrates the flexible validity bit.

In our current implementation, when an append is invoked, we first compute the number of cache lines that an entry spans.
For each group of up to 6 cache lines, a word in the metadata cache line is used for the group.
Each metadata 64-bit word contains 6 validation pairs, each consisting of a 9-bit offset (a cache line spans $512 = 2^9$ bits) and a value bit.
The first metadata word contains also a validity bit (as in the Validity Bit algorithm) to validate the metadata itself.
Since each metadata word uses only 60 bits (6 pairs, each 10 bits), one of the unused bits stores a validity bit.
Figure~\ref{figure:flexible-vb}(b) depicts a log entry.

The code in Listing~\ref{algo-flexible-vb} explains how to write new content to a cache line.
Before writing new content, the algorithm reads the old content and compares it against the new content, in reverse order.
When the xor of an old word and a new word differ, the algorithm counts the number of trailing zeros in the xor'ed value, with the {\sf ctz} instruction (Line~\ref{algo-fvb-ctz}), to calculate the position of the final different bit.
The bit from the new content and its offset is added to the metadata.
Finally, the new content is written in order to ensure that this differentiating bit is written last (Lines~\ref{algo-fvb-copy} -- \ref{algo-fvb-copy2}).

When the old and new data are equal, there is no need to write the new content.
Any bit can serve as the validity bit, so we pick an arbitrary one.

\begin{lstlisting}[label=algo-flexible-vb, caption={Flexible Validity Bit Algorithm},float]
<offset:9, bitValue:1> writeCacheline(word *newcontent, word *log) {
    for j from 7 downto 0 {
        diff = newcontent[j] XOR log[j]
        if (diff != 0) {
            intraWordOffset = ctz(diff); @\label{algo-fvb-ctz}@    // count trailing zeros in the difference word. See Figure@~\ref{figure:flexible-vb}@
            offset = 64 * j + intraWordOffset;
            bitValue = getBit(newcontent[j], intraWordOffset);
            break;
        }
    }
    if (j >= 0) {    // found some difference between old and new content
        for k from 0 upto j - 1  @\label{algo-fvb-copy}@
            log[k]  =newcontent[k];
        atomic_store(&log[j], newcontent[j], memory_order_release); @\label{algo-fvb-copy2}@
        // no need to copy from j+1 to 7. The old and new are the same
    } else {    // all XORs are zeros, the old and new are the same
            // No need to write the new data, just ensure recovery reports this cache line as valid
        offset = 0;    // pick arbitrary offset here (between 0 and 511)
        bitValue = getBit(newcontent[offset / 64], offset % 64);
    }
    return <offset, bitValue>;
}

bool checkCachelineValidity(word *log, offset:9, bitValue:1) { @\label{algo-line-check-fvb}@
    interWordOffset = offset / 64;
    intraWordOffset = offset % 64;
    return getBit(log[interWordOffset], intraWordOffset) == bitValue;
}
\end{lstlisting}

If a program crashes and requires recovery, each cache line is processed as follows.
First, the metadata cache line is validated with its validity bit.
Then, for each group of cache lines, the offset and polarity of its validity bit are compared against the corresponding bit in NVM (the actual data).
If the bits differ, then the new data was not completely written to NVM.
If the bits match, then all bits written before the validity bit were fully written according to the {$\mathcal P$}CSO memory model.
Thus, the cache line is considered as fully written.
Listing~\ref{algo-flexible-vb} also contains the code for the recovery algorithm.

The principle incremental cost of this algorithm is from reading the old content before appending a new entry.
These values, however, can be prefetched from NVM in advance, so that during an append operation, the old value will be in the processor cache.
Additional overhead comes from bitwise xor and computing trailing zeros, but these instructions run quickly on cached values.
Furthermore, it is expected that reading NVM will require significantly lower power than writing to it.

\section{Single Trip Persistent Set}\label{sec-stps}
While logs are an important component for building persistent transactions, it is also possible to directly construct other persistent data structures in NVM using similar techniques.
This section describes a set (hash map) called a Single-Trip Persistent Set (\sset{}) that persists data with a single round-trip to NVM.
In addition to the standard hash map operations, \sset{} supports persistent transactions with all-or-nothing semantics.
The approach used to construct \sset{} can be used for other data structures as well.

The \sset{} design is based on the \vb{} log algorithm\footnote{Designs based on \fvb{} or \random{} are straightforward extension of this algorithm.}.
As with the log algorithm, we do not consider multithreading.

The key idea of the \sset{} algorithm is to store hash entries in a persistent log.
Without failure, the \sset{} algorithm behaves like a standard chaining hash map.
When a failure occurs, however, the hash map can be reconstructed from NVM.

Listing~\ref{listing-logupdate-example} contains the code for the update and recovery procedures.
An Enhanced Persistent Log (Section~\ref{sec:EPL}) is used to store the hash entries durably and it is the only data used during recovery.
The bucket array and next pointers are volatile data that are reconstructed during recovery.
Compared with a normal, volatile hash map, the differences arise from the need to allocate a node, initialize it, and deallocate it in the persistent log.
Searches have no persistence implication and execute normally.

Following a power failure, recovery must be executed before any operation is applied to a \sset{}.
The recovery traverses the entire enhanced log to reconstruct the bucket array and next pointers, so that recovery time is proportional to the amount of memory allocated for the \sset{}. 
The \sset{} algorithm trades fast (and single round-trip) modifications against slow recovery. 
In general, this tradeoff is reasonable since failures are typically rare; the Makalu allocator made a similar concession \cite[c.f. Section~\ref{sec-related}]{Bhandari2016}. 
Set algorithms optimized for fast recovery are outside the scope of this paper. 

\begin{lstlisting}[caption={Hash map updated based on enhanced logging},label={listing-logupdate-example},float]
Global:
    EnhancedPersistentLog hashTableLog;
    node *bucketArray[];
void update(key, data) {
    node **prev, *curr;
    findNode(key, bucketArray, &prev, &curr);    // finds current entry and previous one
    // prev points to either a bucket in bucketArray or next pointer of previous entry
    node *newElement = log.append(key, data);    // returns address of log entry
    *prev = newElement; // links newElement to relevant bucket in bucketArray
    if (curr->key == key) {    // found an older element
        newElement->next = curr->next;    // remove curr from the list
        log.allowReuse(curr);    // free curr to the log
    }
    else {
        newElement->next = curr;
    }
}

void recovery() {
    log.recover();    // bring the log to a consistent state
    for each entry le in hashTableLog from older to newer {
        applyOp(le, bucketArray);    // reapply operation recorded by le
    }
}
\end{lstlisting}

The key difference between the log algorithms discussed preciously and the \sset{} log algorithm (EnhancedPersistentLog) is memory management.
The \sset{} log algorithm is not managed like a queue and it allows entries in the middle of the log to be removed and reused.
This complicates the \vb{} algorithm (also \fvb{}), which assumed it could detect unused entries with a validity bit whose polarity changed only when the log wraps around.
In addition, a conventional log provides a clear ordering among entries, which is required for recovery.
Such an ordering is not available if log entries from the middle of the log are reused.
The enhanced persistent log resolves these issues.

\subsection{Enhanced Persistent Log}
\label{sec:EPL}
The enhanced persistent log is based on \vb{} log algorithm but allows elements to be removed from the middle of the log.
For simplicity, we start with log entries that fit in a single cache line.
Afterward, we consider three extensions: removing elements from the hash map, entries that cross multiple cache lines, and support for transactions.

One issue that removal creates is the order of log entries (which is necessary in recovery to replay actions in the correct order).
In normal logs, entries further from the head are newer.
If elements are removed from the middle, there is no head pointer and the distance between entries is not related to their age.
To solve this issue, we add a version number to each log entry.
The version records the order in which the entries were add to the log.

The second issue is the validity bit in an entry.
For the \vb{} algorithm, 1 initially means \valid{}.
When the log space is reused, the meaning is swapped so that 0 is \valid{}.
This convention is not possible when elements are removed from the middle since adjacent entries may be reused a different number of times, leading to inconsistent polarities.

To solve the validity bit problem, the \sset{} log algorithm uses two validity bits, denoted by $v_0$ and $v_1$.
The entry is \valid{} only if both bits are equal and \invalid{} otherwise.
Listing~\ref{listing-append-enhanced} contains the code to append a new entry to a \sset{}.
Figure~\ref{fig:hash} illustrates the process.

The algorithm starts by finding a {\em reusable} log entry (the specific details of reusability are discussed later).
All log entries are kept in a valid state, so a precondition is that the entry resides in NVM and its two validity bits be equal.
Then the first validity bit $v_0$ is flipped, so it differs from the second bit.
This ensures that if the cache line is flushed early, the validity bits will not match and the entry will be \invalid{}.
At Line~\ref{listing-line-setdata}, the log data is set (in a non-atomic manner).
The release memory fence between flipping the first validity bit and writing the data ensures that the change to the bit reaches the cache before the data.
At Line~\ref{listing-line-incversion}, the version is updated and then the second validity bit is flipped with release semantics, ensuring it reaches the cache after the data update.
Finally, the data is flushed and an \sfence{} is executed, ensuring that the data is stored in NVM.
When append finishes, both validity bits are equal, satisfying the log invariant.

\begin{lstlisting}[caption={Appending data to a \sset{} log},label={listing-append-enhanced},float]
void *append(data) {
    LogEntry *le = log.allocOrReuse();
    assert(both validity bits in le are equal);    // precondition
    bool oldValidity = le->v0;
    le->v0 = !oldValidity; @\label{listing-line-flip1}@
    atomic_thread_fence(memory_order_release);    // ensure flipping happens before writing actual data@\label{listing-line-fence1}@
    for each word w in data:
        le->data[w] = data[w]; @\label{listing-line-setdata}@
    le->ver = versions.increment(); @\label{listing-line-incversion}@
    atomic_store(&le->v1, !oldValidity, memory_order_release); @\label{listing-line-flip2}@
    clflushopt(&le);
    sfence();
}
\end{lstlisting}

To increase performance, the version number and the validity bits can reside in the same machine word (64 bits).
Thus, setting the version and flipping the second validity bit can be done with a single write, which ensures the order between setting the version and the validity bit.

We claim that the \sset{} log algorithm is correct.
\begin{claim}
    Suppose that each log entry fits in a single cache line.
    If a crash occurs after append finished, the recovery procedure would find a log entry with the new data.
    If a crash happens before append finished, the recovery procedure would observe either the log entry with the old data, the log entry with the new data, or an invalid entry.
\end{claim}
\begin{proof}
    After append() finishes, the new data is stored in NVM since is was flushed before the append() operation returns.
    Next consider the case in which the append did not finish, and suppose, by contradiction, that recovery observes a valid entry that consists of a mixture of both old and new data (and not just the old data or the new data).
    Thus, some of the writes at Line~\ref{listing-line-setdata} reached NVM while others did not.
    Let $w1$ be a write that reached NVM and let $w2$ be a write that did not reach NVM.
    Let $flip1$ be the write at Line~\ref{listing-line-flip1} and let $flip2$ be the write at Line~\ref{listing-line-flip2}.
    Due to the release fence at Line~\ref{listing-line-fence1}, we have $flip1 \leq_{hb} w1$.
    Thus according to \cto{}, $flip1 \leq_{p} w1$.
    Due to the release semantics of $flip2$ we have $w2 \leq_{hb} flip2$; according to \cto{}, $w2 \leq_p flip2$.
    Since $w1$ reached NVM, $flip1$ must have reached NVM.
    Since $w2$ did not reach NVM, $flip2$ must not have reached NVM either.
    Thus, the first validity bit must differ from the second bit, contradicting the assumption that the entry was valid.
\end{proof}

\newcommand{\pair}[1]{$<$#1$>$}
\newcommand{\triple}[1]{$<$#1$>$}
\begin{figure}
    \centering
    \includegraphics[height=42mm]{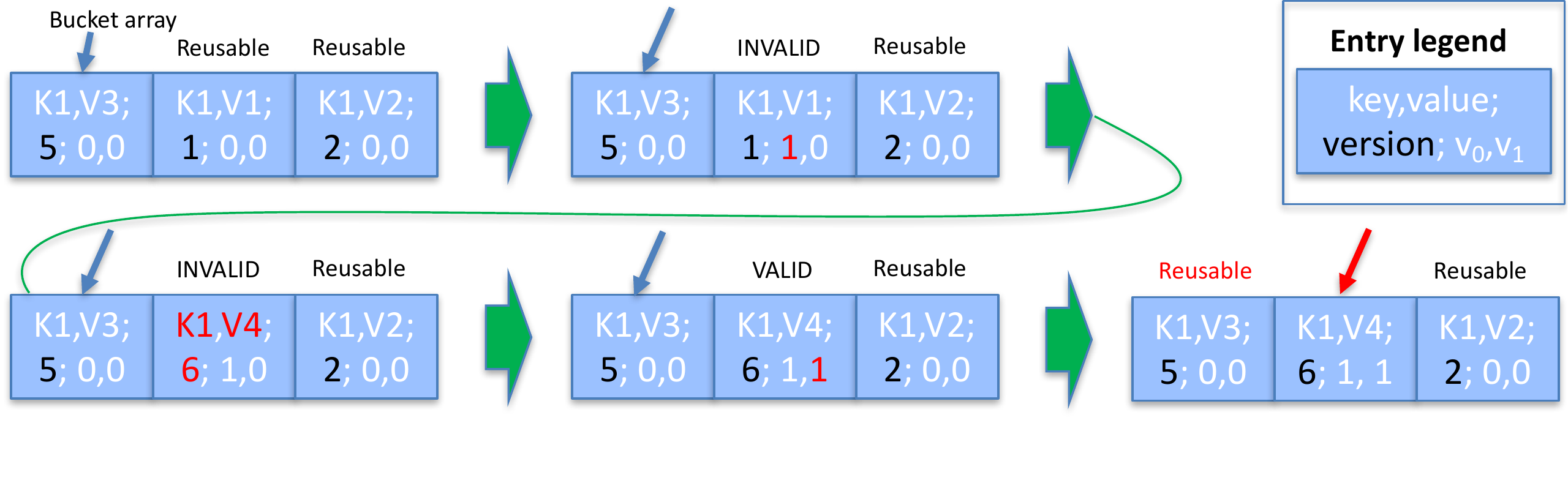}
    \caption{Illustrating update(K1, V4) to a \sset{}.
        Initially K1 is mapped to V3 and there are two reusable entries.
        First, validity bit V0 is flipped to 1, causing the entry to become \invalid{}.
        Then, the key and data are updated to \pair{K1, V4}.
        Afterward, the second validity bit is flipped to 1 and the data is flushed, causing the entry to become \valid{}.
        Finally, the bucketArray is modified to point to the new entry.
}
    \label{fig:hash}
\end{figure}

Next we discuss some more points that are more specific to the hash map implementation.

\subsection{Removing an Element from the Hash Map}
To remove a key from the hash map, a new entry, called a {\em remove entry}, is allocated in the log to specify that the key was removed.
Then, the old entry can be deleted from the log.
However, actually deleting an entry requires a flush operation and is expensive.
To avoid this unnecessary cost, the \sset{} algorithm never {\em delete} entries from the log; it just mark them as possible to reuse.
We call such an entry a {\em reusable entry}.
This means that until the entry is actually reused, it remains valid.
After a power failure, the recovery procedure would treat it as a valid entry
and would insert it into the hash map.
But a subsequent entry with a higher version (such as the remove entry) would remove it from the table.

To avoid unnecessary memory usage, the \sset{} should allow the log also to reuse a remove entry.
This creates a problem if the removed entry is reused before the old entry is reused and a crash occurs, in which case, the recovery procedure would discover the old entry but not the subsequent remove entry, effectively reviving the deleted element.

To prevent this from happening, we used a (volatile) FIFO queue to reuse log entries.
The FIFO ordering ensures that entries are reused in the order in which they are deleted.
Thus, when a remove entry is reused, the corresponding entry has already been overwritten.
During recovery, the reuse FIFO queue must be reconstructed in the correct order.
Alternatively, all entries that are not present in the hash map can be reinitialized, resetting any dependency.

\subsection{Entries with Multiple Cache Lines}
In the case in which an entry spans of more than one cache line, it is possible to extend the \sset{} log algorithm in Listing~\ref{listing-append-enhanced} to operate with two validity bits for each cache line.
An entry is \valid{} if all validity bits are equal and \invalid{} otherwise.
However, frequently, it is possible to use only a single validity bit for cache lines beyond the first.

Consider an old reusable entry consisting of \triple{k1, ver1, val1}, which is replaced by a  new entry \triple{k2, ver2, val2}.
It would be correct for the recovery procedure to observe an entry \triple{k1, ver1, $\alpha$(val1, val2)} where $\alpha$ represents a mixture of its two arguments.
On the other hand, the key and version must be either \pair{k1, ver1} or \pair{k2, ver2}, not be a mixture of both.
If the key and version are \pair{k2, ver2}, the value must be val2.

In general, the key is often smaller than 56 bytes (64 bytes together with the version field).
In this case, only the first cache line needs two validity bits, while other cache lines require only a single validity bit.
If the key is smaller than 119 bytes, it is possible put it entirely in the first two cache lines (plus 8 bytes for the version and validity bits for the first cache line and another byte for the two validity bits for the second cache line).
Thus, it is possible to use 2 validity bits for each of the first two cache lines and one validity bits for each subsequent cache lines.
Listing~\ref{listing-append-multicache} contains code for the first case (56 bytes key).
The only difference from Listing~\ref{listing-append-enhanced} is the setting of multiple validity bits and flushing multiple cache lines at the end of the algorithm and using a release memory fence (instead of a write with release semantics) to ensure that all validity bits are written after all data.

\begin{lstlisting}[caption={Appending data to a \sset{} log spanning multiple cache lines},label={listing-append-multicache},float]
void *append(data) {
    LogEntry *le = findDeletedLE();
    assert(all validity bits in le are equal);    // precondition
    bool oldValidity = le->v0;
    le->v0 = !oldValidity;
    atomic_thread_fence(memory_order_release);    // ensure flipping happens before writing actual data
    for each word w in data:
        le->data[w] = data[w]; 
    le->ver = versions.increment(); 
    atomic_thread_fence(memory_order_release);    // ensure second flipping happens after writing actual data
    le->v1 = !oldValidity;
    for each cacheline cl in le except for the first do:
        le->v[cl] = !oldValidity;    // v[cl] is the single validity bit of cacheline cl
    for each cacheline cl in le do:
        clflushopt(&le->cachelines[cl]);
    sfence();
}
\end{lstlisting}


\subsection{Transactions on a Hash Map}
Since \sset{} uses explicit version numbers, transactions on a \sset{} can be implemented by writing a set of elements with the same version number.
The recovery algorithm must also know how many entries have the same version to decide if a transaction completed successfully or if some elements did not make it to NVM.
In the latter case, all elements of the incomplete transaction must be discarded.
To find the number of elements written in a transaction, we use an 8-bit transactional counter alongside the version number.
When executing a normal operation (without transaction), this counter is set to one.
But when $n$ elements are modified atomically in a transaction, each element has $n$ in its transaction counter and the same version as the other elements.
During recovery, the number of elements with an identical valid version is recorded and compared to the transactional counter.
If they match and each entry is valid, then the transaction is committed.
Otherwise, the transaction did not finish before the failure, and all of its elements are discarded.

\section{Measurements}\label{sec-measurements}
To measure the effectiveness of the log algorithms, we
evaluated their performance on a stress-test micro benchmark, which repeatedly wrote entries to the log.
In addition, we incorporated the new algorithm into two existing systems.
We modified TinySTM, a popular software transaction memory system, to make its transactions persistent.
TinySTM does not have a clear log interface, but it implements transactions with logging.
Second, we used the log algorithm in Atlas \cite{Chakrabarti2014}, a system designed to make
multithreaded applications persistent on machines with NVM memory.
Atlas uses existing locking to delimit persistent atomicity regions.
Atlas already implemented a logging algorithm for NVM, which we replaced with ours.
Finally, we measured the performance of the \sset{} using the YCSB benchmark.

Since NVM components are not commercially available, we followed standard practice and emulated NVM with DRAM.
It is expected that, at least initially, NVM will exhibit higher write latency than read latency.
Thus, following standard practice in this field \cite{Volos2011,Wang2014,Arulraj2015}, we inserted additional delay of 0ns -- 800ns at the sfence operation (which follows clflushopt operations).

The experiments executed on an Intel(R) i7-6700 CPU @ 3.40GHz with 2 Kingston(R) 8GB DDR4 DRAM @ 2133 MHz.
The code was compiled with g++ version 5.4.0.
Unless specified otherwise, each execution was repeated 10 times and the average and 95\% confidence intervals are shown.

\subsection{Log Micro Benchmark}
To measure the effectiveness of the various logging algorithms, we ran a log stress test that repeatedly appended entries to a log.
Every 512 appends, the log entries are read and discarded.
We varied the size of an entry from half a cache line to 1, 2, 4, and 8 full cache lines.
Up to one cache line, we used one metadata word, and for two cache lines and above, we used two metadata words.

We compare 7 log algorithms.
The first was the basic one: write the data and flush it to NVM, then append the new entry to the log by modifying the previous entry's next field.
We consider this variant to be the baseline solution, as it is the simplest to implement and requires no initialization.
However, it requires two round trips to NVM.
This variant is called {\sf \tworounds}.

The second and third variants use checksums to ensure validity.
Both the data and the checksum are written and flush together; an entry is valid if its checksum corresponds to its data.
Dissimilar entries could produce the same checksum, which means that an entry may be reported as valid even though it is only partially written.
This error could result in an arbitrary behavior and opens opportunities for security attacks.

We experimented with cryptographic quality checksum algorithms (e.g., MD5, SHA1), but they are far too expensive for this application.
Instead, we used the CRC64 algorithm\footnote{We used the open source implementation from http://andrewl.dreamhosters.com/filedump}, which offers a low probability of spurious matches at a relatively low cost.
We also used the CRC32 checksum implemented by the x86 crc32 instruction.
This algorithm is weaker and has a greater likelihood of spurious matches.
The CRC32 algorithm is probably not practical for real systems, but it offers a lower performance bound for checksum algorithms.

When a circular buffer is reused, the log contains old entries with a valid checksums.
To avoid reinitializing the log, a checksum should also contain a sequence number that is incremented when the log wraps.
Thus, old entries will have an incorrect checksum with respect to the new sequence number.
It is possible to avoid initializing the log at the beginning, with the (small) risk of arbitrary data being considered valid.

The fourth variant uses the tornbit algorithm (discussed in Section~\ref{sec-related}).
The main drawback of this variant is that it inserts a metadata bit in every word (8 bytes).
Thus, an entry (e.g. string) cannot be accessed directly in a standard way.
With this algorithm, appending to the log and reading from the log require conversions to insert and remove these bits from an entry's representation.

The next are the logging algorithms presented in this paper.
The fifth variant is the validity bit algorithm called \vb{} (Section~\ref{sec:vb}).
It works for up to 2 cache lines.
We do not measure it for larger entries as that requires dividing the payload.

The sixth variant is the first extension to the \vb{} algorithm, which uses random initialization.
During an append it uses one round trip to NVM with probability $1-2^{-64}$ (Section~\ref{sec:random}).
After the log is consumed, it is initialized back to the random value.
These values are not flushed immediately.
Instead, we rely on a subsequent operation to flush the random value, in order to allow initialization to run in the background.
This variant is called \random{}.
The seventh variant uses the flexible validity bit algorithm, called \fvb{} (Section~\ref{sec:fvb}).

\begin{table}
    \centering
    \label{my-label}
    \begin{tabular}{l|lllll}
        Method & Flushes & \begin{tabular}[c]{@{}l@{}}Random\\ incorrect\end{tabular} & \begin{tabular}[c]{@{}l@{}}Adversarial\\ incorrect\end{tabular} & Size limitation & Additional overhead \\\hline
        \vb & n+1 & No & No & $\leq$ 119 bytes* \\
        \fvb & n+1 & No & No & Unlimited & Additional read \\
        \random & 2n+$\frac{n}{2^{64}}$ & No & No & Unlimited &  \\
        Atlas & 1.5n & No & No & 24 bytes &  \\
        tornbit & n+1 & No & No & Unlimited & Slow reads \\
        CRC32 & n & $2^{-32}$ & Yes & Unlimited & Fast (hardware) \\
        CRC64 & n & $2^{-64}$ & Yes & Unlimited & Slow \\
        MD5/SHA1 & n & $\leq 2^{-128}$ & No & Unlimited & Very slow
    \end{tabular}
    \caption{Comparison of logging algorithms.
        The {\em flushes} column is the number of flushes to NVM per entry after the log is reused $n$ times.
        A $+1$ represents initialization; the multiplier of $n$ represents the number of flushes per entry.
        The {\em Random incorrect} column shows the probability that the algorithm is incorrect (and results in an arbitrary behavior after a crash) when the program writes random data.
        {\em Adversarial incorrect} column shows whether an adversarial program can create arbitrary behavior.
        *If possible to interleave data with metadata, then \vb{} has no size limitations.
    }
\label{table-comparison}
\end{table}

All these algorithms require initializing the log before the first use.
The \random{} also performs two writes per append, but the second write is off the append's critical path.

Finally, for Atlas, with entries of size 24 bytes, we used the logging algorithm from Atlas (Section~\ref{sec-related}).
This algorithm adds a next pointer to each log entry and forms a list of all valid entries in the log.
Append starts by writing the data and flushing it to NVM.
Then, the entry is chained to the log by setting the next pointer of the previous element.
If the next pointer of the previous element resides in the same cache line as the entry, a single flush is sufficient.
For 24 bytes entries (and 8 bytes next field), this happens every second entry.
Thus on average, this algorithm requires 1.5 round trips to NVM.
This variant is denoted \atlaslog{}.
Table~\ref{table-comparison} summarizes the properties of all these variants.

\subsubsection{Results}
\begin{figure}
    \centering
    \includegraphics[width=100mm]{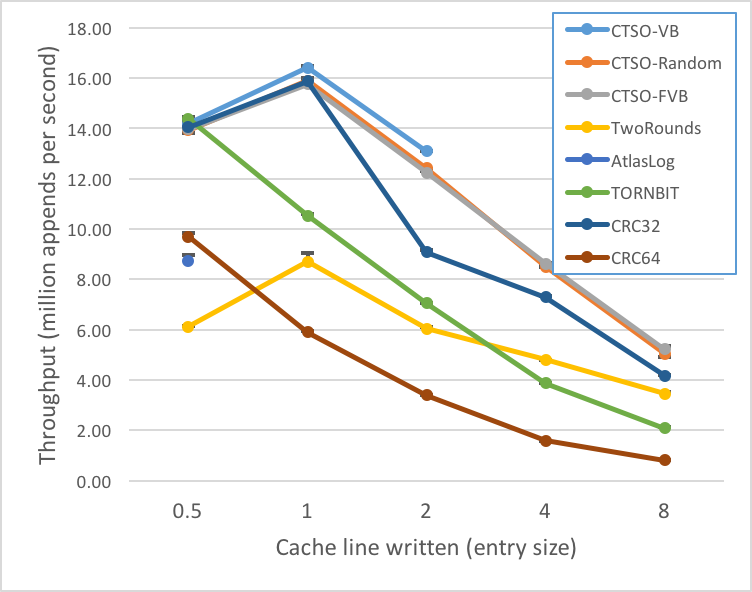} (a)
    \break
    \includegraphics[width=100mm]{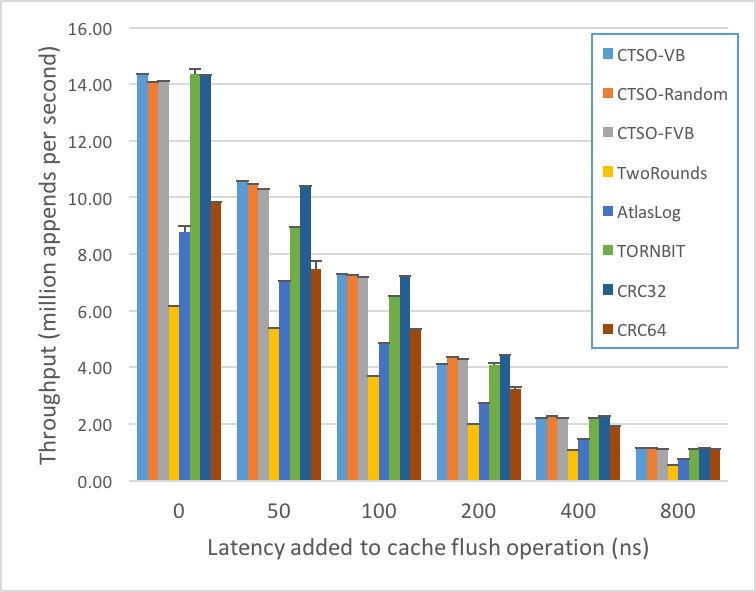} (b)
    \\\noindent
    \caption{Stress testing log algorithm. Throughput (million appends per second) for each algorithm.}
    \label{fig:microLog}
\end{figure}

Figure~\ref{fig:microLog}(a) shows throughput as a function of the number of cache lines written.
When moving from one half to a full cache line, performance increases.
We attribute this to the {\sf clflushopt} instruction, which evicts a line from the processor cache, so that the next operation must fetched it again.
When a full cache line is written, the CPU can prefetch the subsequent cache line, speeding the next reference.
We expect this anomaly to disappear with Intel's proposed {\sf clwb} instruction, which does not evict a line from cache.

The \vb{} algorithm performs best, but works only up to two cache lines (or requires interleaving metadata and data).
The tornbit algorithm is also very efficient for half cache line, but then deteriorates quickly.
This is primarily due to the overhead of reading the log and reconstructing the entry.
The \random{} and \fvb{} algorithm are similar to the \vb{} algorithm, with only 2\% -- 7\% performance loss.
However, they can be extended to larger log entries while leaving the payload consecutive in memory, which \vb{} cannot.
We expect that the \fvb{} algorithm may be a better choice for NVM as it reduces memory wear and may exhibit better power efficiency

The CRC32 algorithm performs close to the \fvb{} and \random{} for one cache line but afterwards is slightly slower.
Recall that CRC32 may erroneously report an entry as valid with non-negligible probability.
The CRC64 has lower performance and for large entry sizes it is even slower than \tworounds.
Finally, \atlaslog{} has better performance than \tworounds, but it is slower than other methods and is restricted to a specific sized entry.
Still, it does not require log initialization.

In Figure~\ref{fig:microLog}(b), we present results with artificial delay added to NVM writes.
The entry size was fixed at a half cache line, so all variants could be compared.
The emulated delay ranges between 0ns and 800ns.
At 800ns, the difference among the variants that flush once is negligible.
As expected, the methods that suffer most from increased latency are those that access NVM more than once, i.e., \tworounds{} and \atlaslog{} (\tworounds{} is slower than \vb{} by 50\% and \atlaslog{} is slower by 33\%).

\subsection{TinySTM}
TinySTM \cite{Felber2008} is a popular software transactional memory system.
To ensures that all writes execute atomically, TinySTM buffers writes during a transaction in a write set and then executes them during the commit.
To make TinySTM persistent, we puts its write entries into a persistent log.
During the commit, all write entries are flushed to NVM to make the transaction persistent.
Then the writes are executed and asynchronously flushed to NVM (i.e., without executing a \sfence{}) since these writes can be recovered from the log).
Periodically, a \sfence{} is executed to flush pending writes and the log is truncated.

The logging mechanism of TinySTM performs variable length appends since the number of write entries is unknown in advance.
We used the \vb{} algorithm since a log entry consists of a set of write entries, each of which is 48 bytes and aligned to a cache line boundary (64 bytes).
We used the 2 extra words as metadata.
When a thread reads an address it wrote in the same transaction, it must read the latest value in the log instead of the older value in memory.
The need to read entries complicates the tornbit algorithm, which does not keep the value in a readable format but rather interleaves it with metadata.
Thus, we did not implement this method.
In addition to the new \vb{} algorithm, we also implemented the log using TwoRounds and CRC64 and CRC32.

\begin{figure}
    \centering
    \includegraphics[width=66.7mm]{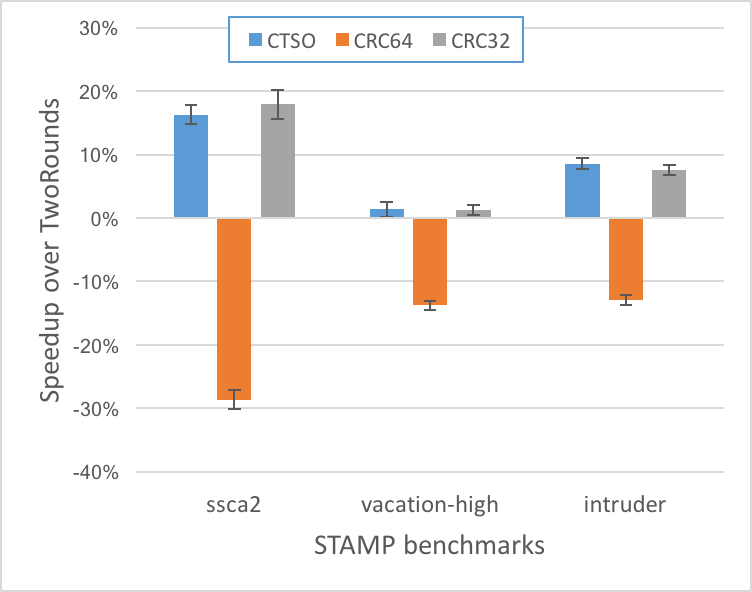}
    \includegraphics[width=65mm]{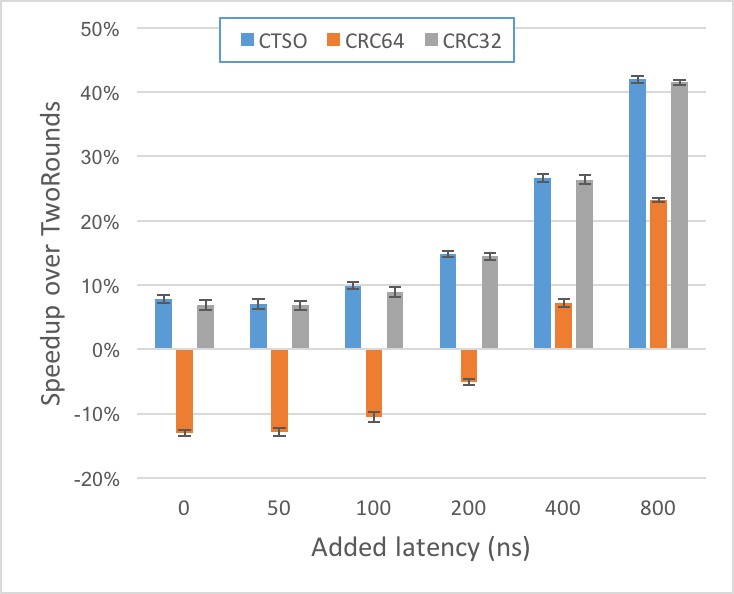}
    \\\noindent
    (a) \hspace{6cm} (b)
    \caption{Speedup over TinySTM using TwoRounds logging algorithm.
    (a) 3 benchmarks from the STAMP benchmark suite. (b) the intruder benchmark from the STAMP suite, varying the added latency of the flush operation between 0 and 800ns.}
    \label{fig:stamp}
\end{figure}

To measure the effectiveness of TinySTM we used 3 benchmarks from the STAMP suite: ssca2, vacation (high contention configuration), and intruder.
For each log algorithm, we created a version of TinySTM that uses the algorithm.
We then ran each STAMP benchmark using each of the TinySTM variants.
We report the ratio between the baseline (TinySTM using {\sf TwoRounds} logging) and the other logging algorithm (higher is better).
The results are depicted in Figure~\ref{fig:stamp}(a).
The CRC64 logging algorithm performs quite poorly in this case, even worse than \tworounds.
By contrast, both the new \vb{} and CRC32 perform better by 1.3\% -- 17.9\%.
But CRC32 offers relatively weak correctness guarantee.

In Figure~\ref{fig:stamp}(b) we introduced artificial latency to the cache flush operation, of between 0 and 800ns and run the intruder benchmark.
As expected, when flushes to NVM are expensive, the benefits of using a single flush to NVM increases.
When the additional latency was 800ns, even CRC64 performs better than the baseline by 23\% while the \vb{} and CRC32 offer 41.5\% -- 42\% improvement in this case.

\subsection{ATLAS}\label{subsection-atlas-results}
Atlas \cite{Chakrabarti2014} is a system designed to simplify porting of existing multithreaded code to run durably on a machine with NVM.
Atlas leverages an application's critical sections (implemented with explicit locks) to delimit internal NVM transactions.
To implement the internal transactions, Atlas uses a specialized NVM logging algorithm.
Each log entry consists of 32 bytes, including 8 bytes pointing to the next entry.
As discussed previously, appending an entry to the log requires 1.5 round trips to NVM on average.

We modified the Atlas system to use the new logging algorithm \vb{} by replacing the next pointer with a validity bit.
In addition to validity, the first element stores the sense of the validity bit for the current iteration (i.e., whether {\em true} means valid or invalid, swapped every reuse).
If a log is exhausted, a new log is allocated.
In this case, the last element in the old log points (via its next pointer) to the first element in the new log.

Atlas also logs {\sf memcpy} and similar functions that write variable length payloads.
In this case, a normal log entry (32 bytes) is added, which points to the variable-length data.
The variable-length entry is written and flushed before the normal log entry is inserted, leading to 2.5 flushes per operation on average.
We replaced this logging algorithm by the \fvb{} algorithm.
The new algorithm also inserts a normal log entry that points to the new data.
However, both the fixed-sized and the variable-sized log entry are written together, so a single round trip to NVM is sufficient.
The flexible validity bit for the first cache line (which contains flexible validity bits for the other cache lines) is stored in the normal log entry that points to the variable length entry.

\begin{figure}
    \centering
    \includegraphics[width=70mm]{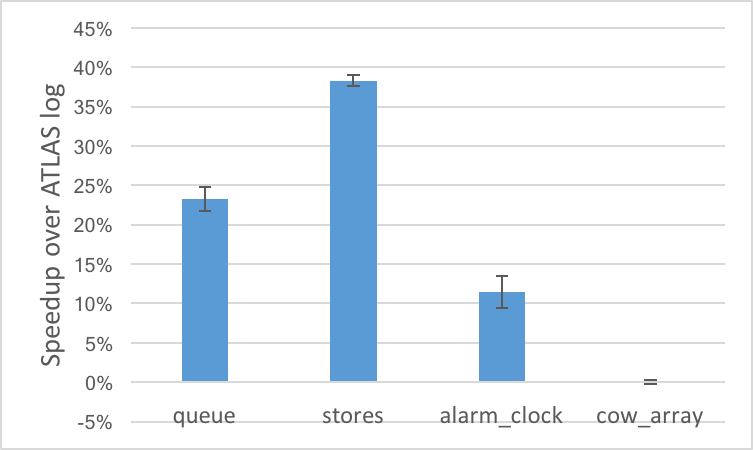}
    \caption{Data structure implemented with Atlas. The ratio between the running time of ATLAS and the running time of Atlas-CSO. Higher is better.}
    \label{fig:atlasds}
\end{figure}

To measure the effectiveness of Atlas with and without the new logging algorithm, we exercised a set of data structures implemented using the Atlas interface and measured their performance.
Four data structures are used: {\em store} modifies an element in a shared array in each iteration, {\em queue} enqueue or dequeue elements from a shared queue, {\em cow\_array} modifies an element in a shared array by creating a new version of the array, and {\em alarm\_clock} modifies an alarm clock while another thread plays the alarm and remove the expired entries.
The original Atlas system is called {\sf ATLAS} while the modified version is called {\sf ATLAS-CSO}.
We present the ratio between the running time of ATLAS and ATLAS-CSO; since these tests are short running, each was executed 100 times to reduce the confidence intervals.
The results appear in Figure~\ref{fig:atlasds}.
The store micro benchmark writes values to an array in a tight loop, so the performance of the log is crucial.
ATLAS-CSO provides a 38\% performance improvement.
For the queue and the alarm clock benchmarks, ATLAS-CSO provides 23\% and 11\% performance improvements, respectively.
The cow-array benchmark uses the log less frequently, so it does not benefit from the change.

\subsection{Single Trip Persistent Set}
\begin{figure}
    \centering
    \includegraphics[width=67mm]{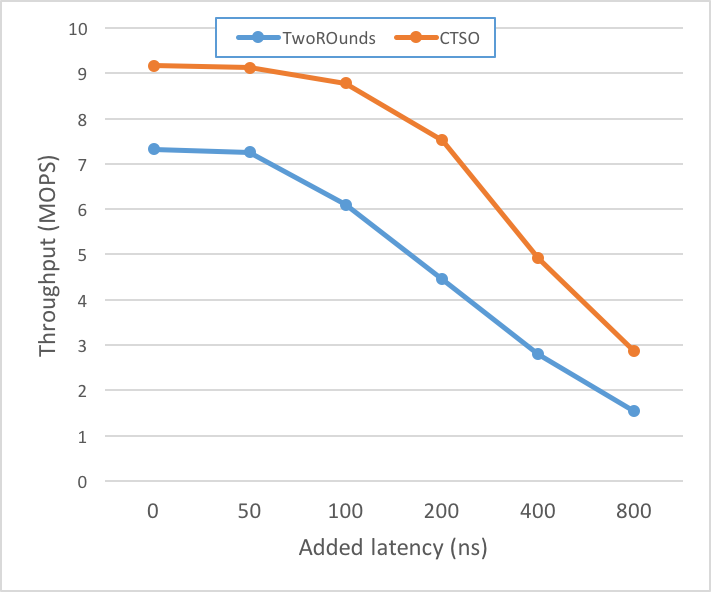}
    \includegraphics[width=67mm]{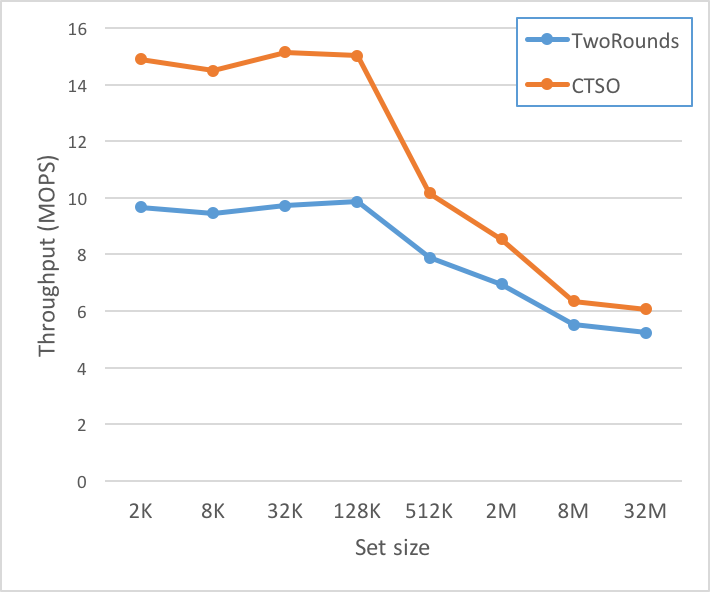}
    \\\noindent
    (a) \hspace{6cm} (b) \\\noindent
    \includegraphics[width=67mm]{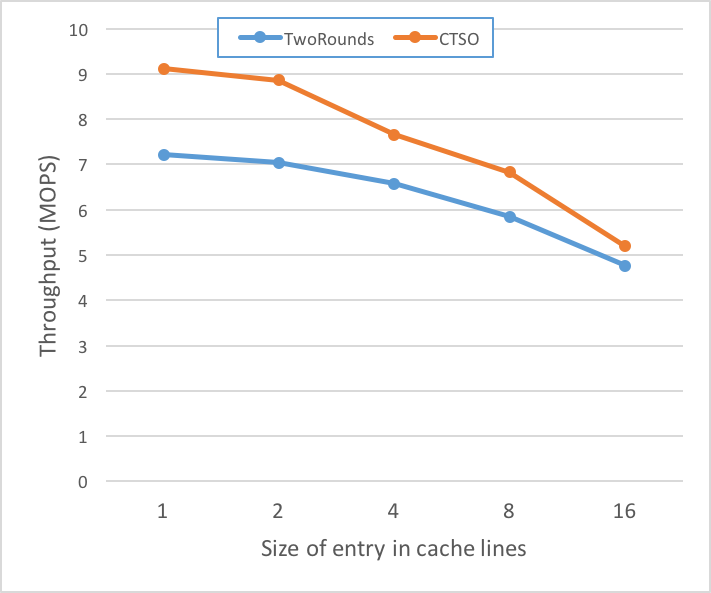}
    \\\noindent (c)
    \caption{Throughput of \sset{} as a function of added flush latency, set size, and entry size.}
    \label{fig:set}
\end{figure}

To measure the effectiveness of the persistent set implementation, we used a stress test with 50\% reads and 50\% updates (based on YCSB workload B).
The \set{} is compared with a baseline implementation that requires two round-trips to persist data (\tworounds{}), one for writing the data and another for writing the next pointer.
As for the log algorithms, we modeled NVM with DRAM and added 0ns -- 800ns delay to cache flush operations to model slower accesses.
Figure~\ref{fig:set}(a) shows that throughput is a function of the additional delay.
With no additional latency (i.e., the latency of NVM is similar to DRAM), \set{} provides 25\% performance improvement over the baseline.
When access to NVM is slower than DRAM, avoiding a second round trip to NVM improves performance up to 86\% over the baseline \tworounds{}.

We also varied the size of the set between 2K and 32M, quadrupling each step.
We did not add additional latency to the flush operation in this case.
The results appear in Figure~\ref{fig:set}(b).
For small sets (up to 128K) the data fits into the processor cache and so the overhead of the flush operations dominants.
Thus, \sset{} improves performance by 52\% -- 56\%.
When the set is larger, navigating the hash map took a significant amount of time, reducing the benefits of \sset{} to 23\% -- 29\%.

Figure~\ref{fig:set}(c) shows the results of varying the size of each node between 1 cache line (64 bytes) and 16 cache lines (1Kb).
The set size was fixed to 1M and no additional latency was added to the flush operation.
\sset{} improves performance 25.9\% -- 26.5\% when entries are one or two cache lines (up to 128 bytes), 16.4\% -- 16.7\% when entries are four or eight cache lines (256 -- 512 bytes), and only 9.1\% when entries consists of 16 cache line (1Kb).

Figure~\ref{fig:set}(b) indicates that for large sets finding the right place to insert the new record is slow, making it desirable to interleave flushing the new record and navigating the hash map.
This is impossible in the baseline implementation since the record's next pointer must be persistent, and it is known only after the record location is found.
However, this is possible in the \sset{} implementation since the new record's next pointer is transient and needs not be flushed to NVM.
This optimization might allow the \sset{} to persist data much faster.

Unfortunately, this optimization actually {\em reduces} performance in current architectures.
This is likely due to the {\sf clflushopt} instruction, which evicts the cache line from the processor cache.
If the entry is flushed before navigating the data structure, then the cache line containing the new record is evicted from cache.
When the record's next pointer is set, the processor must bring the cache line back from memory.
We expect this inefficiency to disappear once the {\sf clwb} instruction becomes available, as it writes a cache line to NVM without evicting it from the processor cache.
Code for this optimization appears in Appendix~\ref{appendix-optimize-sset}.

\section{Related Work}\label{sec-related}
Mnemosyne is a system for low-level management of persistent memory \cite{Volos2011}.
It defined new hardware primitives and programming language type annotations for persistent regions and implemented a persistent log and transactions.
Their log algorithm used the tornbit optimization: data is broken so that each word has a 63 bits payload and one bit metadata.
The metadata is a validity bit and allows recovery to determine whether the word was written or not.
The main overhead of Mnemosyne is reading the log; it is not possible to access data in the log directly, but instead its contents must be reassembled by removing metadata bits. 
This paper presents several algorithms that perform as well or better than tornbit, but do not modify a log entry's representation.

Several systems (\cite{Hu2017}, \cite{Kolli2016} based on \citet{Prabhakaran2005}) used checksums to validate the integrity of log appends.
\citet{Kolli2016} also studied hardware support for transactions to provide NVM atomicity and interleaved flushes to NVM with subsequent flushes, to reduce the length of the critical path of flushes.
\citet{Hu2017} designed log structured non-volatile memory management system. Their system turns all writes into
NVM to log appends. The log is indexed in volatile memory so that data can be located efficiently.
This paper presents several logging algorithms that perform as well as checksumming, but do not have the same (small) margin of error and are not vulnerable to collision attacks.

DudeTM \cite{Liu2017a} is a recent transactional system for NVM.
It reduces the overhead of logging by persisting the log in batch in the background.
This increases the latency of persisting the data, which is lost if a power-failure happens before the background thread finishes.
The logging algorithms in this paper ensure that data is stored non-volatility before the return.
They may also be used to speed logging performed by a background thread.

The systems described above assume that all writes to NVM are part of a transaction.
Thus, every write to NVM must be logged.
This emphasizes the importance of the log for the overall performance of the system.

Atlas \cite{Chakrabarti2014} is a system that uses explicit locks to delimit regions of code whose outputs should be persisted atomically in NVM.
Locks differ from transactions since partial overlapping of locks creates non-trivial atomicity regions that have no equivalent with transactions.
Atlas ensures atomicity through logging.
Each log entry is 32 bytes, including a next pointer.
Ensuring durability requires two phases: first, the new entry is flushed and then, the previous entry's next pointer is modified to point to the new node.
But since every other log entry resides in the same cache line, one of two times, a single flush suffices.
This paper explores only logging, and, as shown in Section~\ref{subsection-atlas-results}, its algorithms can improve the performance of Atlas by avoiding the second round trip.
Atlas uses log-elision to avoid logging writes to NVM outside of critical sections.
Writes inside a critical sections must still be logged.


There are also hardware solutions for NVM logging.
\citet{Joshi2017} incorporated logging into the memory controller.
It is responsible for respecting ordering constraints, but it is off the processor's critical path.
The hardware implements ``undo logging'' by recording the old value of each modified cache line.
Another hardware solution was presented by \citet{Doshi2016a}.
Their system implements ``redo logging'' but avoids the overhead on reads by not flushing the cache to the NVM before the log is written.
When needed, they used a victim cache to store the data.
Other proposals (\cite{EpochB,Pelley2014, Lu2014}) augment the cache subsystem to flush cache lines to NVM in a desired order.
The algorithms in this paper work with existing processors.

A commonly used solution to reduce the overhead of appending is batching \cite{Huang2014,Arulraj2015,Pelley2013}.
Instead of flushing data immediately, modifications to multiple elements are accumulated and flushed to NVM together.
Batching reduces the durability guarantee of the log.
In addition, there are cases in which batching should not be used since important data may be overwritten before the log appears in NVM.
This paper provides a strict durability guarantee: once the operation returns, the data is persistent in NVM.
Of course, there are performance benefits to batching that might make sense in particular circumstances, and this optimization could be added to the algorithms in this paper.

A persistent log is a crucial building block for constructing persistent transactions.
However, it is also possible to implement transactions and persistent data structures with copy-on-write.
Copy-on-write first builds a new state and then replaces the existing state with the new with a single atomic operation, usually a single 8-bytes modification.
This paradigm is popular for hard-disks and has also considered for NVM \cite{Venkataraman2011,Schwalb,EpochB}.
\citet{Arulraj2015} compared ``copy-on-write'' to modification-in-place via logging for NVM.
They reported that the log solution was a better alternative.
Copy-on-write algorithms require at least two round trips to NVM: one to write the new state and another to switch the current state with the new state.
This paper proposes several algorithms that only require a single round trip to NVM and shows their performance benefits.

\citet{Bhandari2016} proposes a memory allocator for NVM called Makalu that avoids cache flush operations.
Their algorithm does not flush data to NVM for standard allocation and free.
After a failure, it restores the allocator to a consistent state by performing a garbage collection cycle on the persistent heap to reclaim lost objects.
The single trip persistent set (\sset{}) algorithm in this paper also avoids flushing when reusing memory locations, but it does not eliminate flushes in general.
Moreover, the \sset{} algorithm offers a key-value store interface, while Makalu provides an allocator interface that requires additional mechanisms to find and store values. 
Makalu relies on garbage collection to restore the allocator state after a failure, while \sset{} uses the semantics of a key-value store to identify and free entries that do not contain current data. 

\citet{Boehm2016} studied persistent programming models for NVM.
They consider the trade off between restrictions imposed on the programmer and the need to log updates outside of critical sections.
Their findings apply to this work as well, but the memory model and algorithms in this paper are aimed at the system designer and are not expected to be directly visible to a programmer.

\section{Conclusion}\label{sec-conclusion}
This paper presents the persistent cache store order ($\mathcal P$CSO).
We start with a model of the memory system in which writes to different cache lines are not ordered,
but writes to the same cache line occur in the order in which they become visible to other threads.
From this, we build a logging algorithm that ensures atomicity of appending to the log while requiring only a single round trip (flush) to non-volatile memory.
The basic algorithm (\vb{}) uses a single bit per cache line to ensure appends are written atomically.
We also discuss two extensions, \random{} and \fvb{}, that persist larger amounts of data.
The new algorithms provide significant performance improvement over prior logging algorithms.
Incorporating the \vb{} algorithm into TinySTM and Atlas produces performance improvement up to 16\% and 12\%, respectively.
In general, the \vb{} and \fvb{} algorithms perform well, but for specific log entry sizes, other techniques may be worth considering.

Finally, we extend the logging algorithm to a persistent set (hashmap).
This set also requires only a single round trip to NVM for its modification operation.
Furthermore, it also provides transactions (with a limited amount of modifications) at the data structure level.

\section*{Acknowledgment}
The authors would like to thank the anonymous reviewers of OOPSLA for their help on improving the paper and particularly Section~\ref{sec-memoryorder} and Section~\ref{sec-stps}.

\appendix
\section{Optimizing \set{}}\label{appendix-optimize-sset}
\begin{lstlisting}[label=algo-optimize-sset, caption={\sset{} optimization}]
struct metadata{
    long_1b v1:1;
    long_1b v2:1;
    long_8b txncount:8;
    long_54b version:54;
}
    node **prev, *curr;
findNode(key, bucketArray, &prev, &curr);    // finds current entry and previous one

void update(k, v) {
    node *newentry  = log.reuseOrAlloc();
    bool localV1 = newentry->meta.flipV1();
    atomic_thread_fence(memory_order_release);    // ensure flipping v1 happens before other writes
    newentry->k = k;
    newentry->v = v;
    long_54b curr_ver = versions.increment();
    // Ensure v2 is written last by store with release semantics
    newentry->meta.set(ver = curr_ver, txncount = 1, v2 = localV1, memory_order_release);
    clflushopt(newentry);    // flush cache line asynchroneously

    node **prev, *curr; {\tiny }
    findNode(key, bucketArray, &prev, &curr); // navigate while cache line is flushed
    *prev = newElement; // links newElement to relevant bucket in bucketArray
    if (curr->k == k) {    // k exists, replace curr with newentry
        newentry->next = next->next;
        log.allowReuse(curr);
    }
    else {    // inserting new k,v pair between prev and curr. 
        newentry->next = curr;
    }
    // current version flushed newentry here, after the newentry->next is set
    sfence();    // ensure that clflushopt(newentry) is finished before we return from operation
}
\end{lstlisting}

\bibliography{nvm,MemoryModel}


\begin{thebibliography}{00}


\ifx \showCODEN    \undefined \def \showCODEN     #1{\unskip}     \fi
\ifx \showDOI      \undefined \def \showDOI       #1{{\tt DOI:}\penalty0{#1}\ }
  \fi
\ifx \showISBNx    \undefined \def \showISBNx     #1{\unskip}     \fi
\ifx \showISBNxiii \undefined \def \showISBNxiii  #1{\unskip}     \fi
\ifx \showISSN     \undefined \def \showISSN      #1{\unskip}     \fi
\ifx \showLCCN     \undefined \def \showLCCN      #1{\unskip}     \fi
\ifx \shownote     \undefined \def \shownote      #1{#1}          \fi
\ifx \showarticletitle \undefined \def \showarticletitle #1{#1}   \fi
\ifx \showURL      \undefined \def \showURL       #1{#1}          \fi
\providecommand\bibfield[2]{#2}
\providecommand\bibinfo[2]{#2}
\providecommand\natexlab[1]{#1}
\providecommand\showeprint[2][]{arXiv:#2}

\bibitem[\protect\citeauthoryear{Akinaga and Shima}{Akinaga and Shima}{2010}]%
        {Akinaga2010}
\bibfield{author}{\bibinfo{person}{Hiroyuki Akinaga} {and}
  \bibinfo{person}{Hisashi Shima}.} \bibinfo{year}{2010}\natexlab{}.
\newblock \showarticletitle{{Resistive Random Access Memory (ReRAM) Based on
  Metal Oxides}}.
\newblock \bibinfo{journal}{{\em Proc. IEEE\/}} \bibinfo{volume}{98},
  \bibinfo{number}{12} (\bibinfo{date}{dec} \bibinfo{year}{2010}),
  \bibinfo{pages}{2237--2251}.
\newblock
\showDOI{%
\url{http://dx.doi.org/10.1109/JPROC.2010.2070830}}


\bibitem[\protect\citeauthoryear{Arulraj and Pavlo}{Arulraj and Pavlo}{2015}]%
        {Arulraj2015}
\bibfield{author}{\bibinfo{person}{Joy Arulraj} {and} \bibinfo{person}{Andrew
  Pavlo}.} \bibinfo{year}{2015}\natexlab{}.
\newblock \showarticletitle{{Let's Talk About Storage {\&} Recovery Methods for
  Non-Volatile Memory Database Systems}}.
\newblock \bibinfo{journal}{{\em Proc. 2015 ACM SIGMOD Int. Conf. Manag.
  Data\/}} \bibinfo{number}{1} (\bibinfo{year}{2015}),
  \bibinfo{pages}{707--722}.
\newblock
\showISBNx{9781450327589}
\showISSN{07308078}
\showDOI{%
\url{http://dx.doi.org/10.1145/2723372.2749441}}


\bibitem[\protect\citeauthoryear{Avni and Brown}{Avni and Brown}{2016}]%
        {Avni2016}
\bibfield{author}{\bibinfo{person}{Hillel Avni} {and} \bibinfo{person}{Trevor
  Brown}.} \bibinfo{year}{2016}\natexlab{}.
\newblock \showarticletitle{{PHyTM: Persistent Hybrid Transactional Memory.}}
\newblock \bibinfo{journal}{{\em PVLDB\/}} \bibinfo{volume}{10},
  \bibinfo{number}{4} (\bibinfo{year}{2016}), \bibinfo{pages}{409--420}.
\newblock
\showURL{%
\url{http://www.vldb.org/pvldb/vol10/p409-brown.pdf}}


\bibitem[\protect\citeauthoryear{Bhandari, Chakrabarti, and Boehm}{Bhandari
  et~al\mbox{.}}{2016}]%
        {Bhandari2016}
\bibfield{author}{\bibinfo{person}{Kumud Bhandari}, \bibinfo{person}{Dhruva~R.
  Chakrabarti}, {and} \bibinfo{person}{Hans-J. Boehm}.}
  \bibinfo{year}{2016}\natexlab{}.
\newblock \showarticletitle{{Makalu: fast recoverable allocation of
  non-volatile memory}}.
\newblock \bibinfo{journal}{{\em Proc. 2016 ACM SIGPLAN Int. Conf.
  Object-Oriented Program. Syst. Lang. Appl. - OOPSLA 2016\/}}
  (\bibinfo{year}{2016}), \bibinfo{pages}{677--694}.
\newblock
\showISBNx{9781450344449}
\showDOI{%
\url{http://dx.doi.org/10.1145/2983990.2984019}}


\bibitem[\protect\citeauthoryear{Boehm, Adve, Boehm, and Adve}{Boehm
  et~al\mbox{.}}{2008}]%
        {Boehm2008}
\bibfield{author}{\bibinfo{person}{Hans-J. Boehm}, \bibinfo{person}{Sarita~V.
  Adve}, \bibinfo{person}{Hans-J. Boehm}, {and} \bibinfo{person}{Sarita~V.
  Adve}.} \bibinfo{year}{2008}\natexlab{}.
\newblock \showarticletitle{{Foundations of the C++ concurrency memory model}}.
  In \bibinfo{booktitle}{{\em Proc. 2008 ACM SIGPLAN Conf. Program. Lang. Des.
  Implement. - PLDI '08}}, Vol.~\bibinfo{volume}{43}. \bibinfo{publisher}{ACM
  Press}, \bibinfo{address}{New York, New York, USA}, \bibinfo{pages}{68}.
\newblock
\showISBNx{9781595938602}
\showISSN{0362-1340}
\showDOI{%
\url{http://dx.doi.org/10.1145/1375581.1375591}}


\bibitem[\protect\citeauthoryear{Boehm and Chakrabarti}{Boehm and
  Chakrabarti}{2016}]%
        {Boehm2016}
\bibfield{author}{\bibinfo{person}{Hans-J. Boehm} {and}
  \bibinfo{person}{Dhruva~R. Chakrabarti}.} \bibinfo{year}{2016}\natexlab{}.
\newblock \showarticletitle{{Persistence programming models for non-volatile
  memory}}.
\newblock \bibinfo{journal}{{\em Proc. 2016 ACM SIGPLAN Int. Symp. Mem. Manag.
  - ISMM 2016\/}} (\bibinfo{year}{2016}), \bibinfo{pages}{55--67}.
\newblock
\showISBNx{9781450343176}
\showISSN{13686798}
\showDOI{%
\url{http://dx.doi.org/10.1145/2926697.2926704}}


\bibitem[\protect\citeauthoryear{Chakrabarti, Boehm, and Bhandari}{Chakrabarti
  et~al\mbox{.}}{2014}]%
        {Chakrabarti2014}
\bibfield{author}{\bibinfo{person}{Dhruva~R Chakrabarti},
  \bibinfo{person}{Hans-J. Boehm}, {and} \bibinfo{person}{Kumud Bhandari}.}
  \bibinfo{year}{2014}\natexlab{}.
\newblock \showarticletitle{{Atlas: Leveraging Locks for Non-volatile Memory
  Consistency}}.
\newblock \bibinfo{journal}{{\em Proc. 2014 ACM Int. Conf. Object Oriented
  Program. Syst. Lang. {\&}{\#}38; Appl.\/}} (\bibinfo{year}{2014}),
  \bibinfo{pages}{433--452}.
\newblock
\showISBNx{978-1-4503-2585-1}
\showISSN{15232867}
\showDOI{%
\url{http://dx.doi.org/10.1145/2660193.2660224}}
\showeprint[arxiv]{2660224}


\bibitem[\protect\citeauthoryear{Condit, Nightingale, Frost, Ipek, Lee, Burger,
  and Coetzee}{Condit et~al\mbox{.}}{2009}]%
        {EpochB}
\bibfield{author}{\bibinfo{person}{Jeremy Condit}, \bibinfo{person}{Edmund~B.
  Nightingale}, \bibinfo{person}{Christopher Frost}, \bibinfo{person}{Engin
  Ipek}, \bibinfo{person}{Benjamin Lee}, \bibinfo{person}{Doug Burger}, {and}
  \bibinfo{person}{Derrick Coetzee}.} \bibinfo{year}{2009}\natexlab{}.
\newblock \showarticletitle{{Better I/O through byte-addressable, persistent
  memory}}. In \bibinfo{booktitle}{{\em Proc. ACM SIGOPS 22nd Symp. Oper. Syst.
  Princ. - SOSP '09}}. \bibinfo{publisher}{ACM Press}, \bibinfo{address}{New
  York, New York, USA}, \bibinfo{pages}{133}.
\newblock
\showISBNx{9781605587523}
\showDOI{%
\url{http://dx.doi.org/10.1145/1629575.1629589}}


\bibitem[\protect\citeauthoryear{Doshi, Giles, and Varman}{Doshi
  et~al\mbox{.}}{2016}]%
        {Doshi2016a}
\bibfield{author}{\bibinfo{person}{Kshitij Doshi}, \bibinfo{person}{Ellis
  Giles}, {and} \bibinfo{person}{Peter Varman}.}
  \bibinfo{year}{2016}\natexlab{}.
\newblock \showarticletitle{{Atomic persistence for SCM with a non-intrusive
  backend controller}}. In \bibinfo{booktitle}{{\em 2016 IEEE Int. Symp. High
  Perform. Comput. Archit.}} \bibinfo{publisher}{IEEE},
  \bibinfo{pages}{77--89}.
\newblock
\showISBNx{978-1-4673-9211-2}
\showDOI{%
\url{http://dx.doi.org/10.1109/HPCA.2016.7446055}}


\bibitem[\protect\citeauthoryear{Felber, Fetzer, and Riegel}{Felber
  et~al\mbox{.}}{2008}]%
        {Felber2008}
\bibfield{author}{\bibinfo{person}{Pascal Felber}, \bibinfo{person}{Christof
  Fetzer}, {and} \bibinfo{person}{Torvald Riegel}.}
  \bibinfo{year}{2008}\natexlab{}.
\newblock \showarticletitle{{Dynamic performance tuning of word-based software
  transactional memory}}. In \bibinfo{booktitle}{{\em Proc. 13th ACM SIGPLAN
  Symp. Princ. Pract. parallel Program. - PPoPP '08}}. \bibinfo{publisher}{ACM
  Press}, \bibinfo{address}{New York, New York, USA}, \bibinfo{pages}{237}.
\newblock
\showISBNx{9781595937957}
\showDOI{%
\url{http://dx.doi.org/10.1145/1345206.1345241}}


\bibitem[\protect\citeauthoryear{Hosomi, Yamagishi, Yamamoto, Bessho, Higo,
  Yamane, Yamada, Shoji, Hachino, Fukumoto, Nagao, and Kano}{Hosomi
  et~al\mbox{.}}{2005}]%
        {Hosomi}
\bibfield{author}{\bibinfo{person}{M. Hosomi}, \bibinfo{person}{H. Yamagishi},
  \bibinfo{person}{T. Yamamoto}, \bibinfo{person}{K. Bessho},
  \bibinfo{person}{Y. Higo}, \bibinfo{person}{K. Yamane}, \bibinfo{person}{H.
  Yamada}, \bibinfo{person}{M. Shoji}, \bibinfo{person}{H. Hachino},
  \bibinfo{person}{C. Fukumoto}, \bibinfo{person}{H. Nagao}, {and}
  \bibinfo{person}{H. Kano}.} \bibinfo{year}{2005}\natexlab{}.
\newblock \showarticletitle{{A novel nonvolatile memory with spin torque
  transfer magnetization switching: spin-ram}}. In \bibinfo{booktitle}{{\em
  IEEE Int. Devices Meet. 2005. IEDM Tech. Dig.}} \bibinfo{publisher}{IEEE},
  \bibinfo{pages}{459--462}.
\newblock
\showISBNx{0-7803-9268-X}
\showDOI{%
\url{http://dx.doi.org/10.1109/IEDM.2005.1609379}}


\bibitem[\protect\citeauthoryear{Hu, Ren, Badam, and Moscibroda}{Hu
  et~al\mbox{.}}{2017}]%
        {Hu2017}
\bibfield{author}{\bibinfo{person}{Qingda Hu}, \bibinfo{person}{Jinglei Ren},
  \bibinfo{person}{Anirudh Badam}, {and} \bibinfo{person}{Thomas Moscibroda}.}
  \bibinfo{year}{2017}\natexlab{}.
\newblock \showarticletitle{{Log-Structured Non-Volatile Main Memory}}. In
  \bibinfo{booktitle}{{\em 2017 USENIX Annu. Tech. Conf. (USENIX ATC 17)}}.
\newblock
\showURL{%
\url{http://jinglei.ren.systems/lsnvmm}}


\bibitem[\protect\citeauthoryear{Huang, Schwan, and Qureshi}{Huang
  et~al\mbox{.}}{2014}]%
        {Huang2014}
\bibfield{author}{\bibinfo{person}{Jian Huang}, \bibinfo{person}{K Schwan},
  {and} \bibinfo{person}{Mk Qureshi}.} \bibinfo{year}{2014}\natexlab{}.
\newblock \showarticletitle{{NVRAM-aware Logging in Transaction Systems}}.
\newblock \bibinfo{journal}{{\em Proc. VLDB Endow.\/}} \bibinfo{volume}{8},
  \bibinfo{number}{4} (\bibinfo{year}{2014}), \bibinfo{pages}{389--400}.
\newblock
\showISSN{21508097}
\showDOI{%
\url{http://dx.doi.org/10.14778/2735496.2735502}}


\bibitem[\protect\citeauthoryear{Joshi, Nagarajan, Viglas, and Cintra}{Joshi
  et~al\mbox{.}}{2017}]%
        {Joshi2017}
\bibfield{author}{\bibinfo{person}{A Joshi}, \bibinfo{person}{V Nagarajan},
  \bibinfo{person}{S Viglas}, {and} \bibinfo{person}{M Cintra}.}
  \bibinfo{year}{2017}\natexlab{}.
\newblock \showarticletitle{{ATOM: Atomic Durability in Non-volatile Memory
  through Hardware Logging}}.
\newblock \bibinfo{journal}{{\em 23rd IEEE Symp. High Perform. Comput. Archit.
  - HPCA'17\/}} (\bibinfo{year}{2017}).
\newblock


\bibitem[\protect\citeauthoryear{Kolli, Pelley, Saidi, Chen, and Wenisch}{Kolli
  et~al\mbox{.}}{2016}]%
        {Kolli2016}
\bibfield{author}{\bibinfo{person}{Aasheesh Kolli}, \bibinfo{person}{Steven
  Pelley}, \bibinfo{person}{Ali Saidi}, \bibinfo{person}{Peter~M Chen}, {and}
  \bibinfo{person}{Thomas~F Wenisch}.} \bibinfo{year}{2016}\natexlab{}.
\newblock \showarticletitle{{High-Performance Transactions for Persistent
  Memories}}.
\newblock \bibinfo{journal}{{\em Asplos\/}} (\bibinfo{year}{2016}),
  \bibinfo{pages}{399--411}.
\newblock
\showISBNx{9781450340915}
\showISSN{03621340}
\showDOI{%
\url{http://dx.doi.org/10.1145/2872362.2872381}}


\bibitem[\protect\citeauthoryear{Lee, Ipek, Mutlu, Burger, Lee, Ipek, Mutlu,
  and Burger}{Lee et~al\mbox{.}}{2009}]%
        {Lee2009}
\bibfield{author}{\bibinfo{person}{Benjamin~C. Lee}, \bibinfo{person}{Engin
  Ipek}, \bibinfo{person}{Onur Mutlu}, \bibinfo{person}{Doug Burger},
  \bibinfo{person}{Benjamin~C. Lee}, \bibinfo{person}{Engin Ipek},
  \bibinfo{person}{Onur Mutlu}, {and} \bibinfo{person}{Doug Burger}.}
  \bibinfo{year}{2009}\natexlab{}.
\newblock \showarticletitle{{Architecting phase change memory as a scalable
  dram alternative}}. In \bibinfo{booktitle}{{\em Proc. 36th Annu. Int. Symp.
  Comput. Archit. - ISCA '09}}, Vol.~\bibinfo{volume}{37}.
  \bibinfo{publisher}{ACM Press}, \bibinfo{address}{New York, New York, USA},
  \bibinfo{pages}{2}.
\newblock
\showISBNx{9781605585260}
\showDOI{%
\url{http://dx.doi.org/10.1145/1555754.1555758}}


\bibitem[\protect\citeauthoryear{Liu, Zhang, Chen, Qian, Wu, Zheng, Ren, Liu,
  Zhang, Chen, Qian, Wu, Zheng, Ren, Liu, Zhang, Chen, Qian, Wu, Zheng, Ren,
  Liu, Zhang, Chen, Qian, Wu, Zheng, and Ren}{Liu et~al\mbox{.}}{2017}]%
        {Liu2017a}
\bibfield{author}{\bibinfo{person}{Mengxing Liu}, \bibinfo{person}{Mingxing
  Zhang}, \bibinfo{person}{Kang Chen}, \bibinfo{person}{Xuehai Qian},
  \bibinfo{person}{Yongwei Wu}, \bibinfo{person}{Weimin Zheng},
  \bibinfo{person}{Jinglei Ren}, \bibinfo{person}{Mengxing Liu},
  \bibinfo{person}{Mingxing Zhang}, \bibinfo{person}{Kang Chen},
  \bibinfo{person}{Xuehai Qian}, \bibinfo{person}{Yongwei Wu},
  \bibinfo{person}{Weimin Zheng}, \bibinfo{person}{Jinglei Ren},
  \bibinfo{person}{Mengxing Liu}, \bibinfo{person}{Mingxing Zhang},
  \bibinfo{person}{Kang Chen}, \bibinfo{person}{Xuehai Qian},
  \bibinfo{person}{Yongwei Wu}, \bibinfo{person}{Weimin Zheng},
  \bibinfo{person}{Jinglei Ren}, \bibinfo{person}{Mengxing Liu},
  \bibinfo{person}{Mingxing Zhang}, \bibinfo{person}{Kang Chen},
  \bibinfo{person}{Xuehai Qian}, \bibinfo{person}{Yongwei Wu},
  \bibinfo{person}{Weimin Zheng}, {and} \bibinfo{person}{Jinglei Ren}.}
  \bibinfo{year}{2017}\natexlab{}.
\newblock \showarticletitle{{DudeTM: Building Durable Transactions with
  Decoupling for Persistent Memory}}. In \bibinfo{booktitle}{{\em Proc.
  Twenty-Second Int. Conf. Archit. Support Program. Lang. Oper. Syst. - ASPLOS
  '17}}, Vol.~\bibinfo{volume}{45}. \bibinfo{publisher}{ACM Press},
  \bibinfo{address}{New York, New York, USA}, \bibinfo{pages}{329--343}.
\newblock
\showISBNx{9781450344654}
\showDOI{%
\url{http://dx.doi.org/10.1145/3037697.3037714}}


\bibitem[\protect\citeauthoryear{Lu, Shu, Sun, and Mutlu}{Lu
  et~al\mbox{.}}{2014}]%
        {Lu2014}
\bibfield{author}{\bibinfo{person}{Youyou Lu}, \bibinfo{person}{Jiwu Shu},
  \bibinfo{person}{Long Sun}, {and} \bibinfo{person}{Onur Mutlu}.}
  \bibinfo{year}{2014}\natexlab{}.
\newblock \showarticletitle{{Loose-Ordering Consistency for persistent
  memory}}. In \bibinfo{booktitle}{{\em 2014 IEEE 32nd Int. Conf. Comput.
  Des.}} \bibinfo{publisher}{IEEE}, \bibinfo{pages}{216--223}.
\newblock
\showISBNx{978-1-4799-6492-5}
\showDOI{%
\url{http://dx.doi.org/10.1109/ICCD.2014.6974684}}


\bibitem[\protect\citeauthoryear{Manson, Pugh, Adve, Manson, Pugh, and
  Adve}{Manson et~al\mbox{.}}{2005}]%
        {Manson2005}
\bibfield{author}{\bibinfo{person}{Jeremy Manson}, \bibinfo{person}{William
  Pugh}, \bibinfo{person}{Sarita~V. Adve}, \bibinfo{person}{Jeremy Manson},
  \bibinfo{person}{William Pugh}, {and} \bibinfo{person}{Sarita~V. Adve}.}
  \bibinfo{year}{2005}\natexlab{}.
\newblock \showarticletitle{{The Java memory model}}. In
  \bibinfo{booktitle}{{\em Proc. 32nd ACM SIGPLAN-SIGACT sysposium Princ.
  Program. Lang. - POPL '05}}, Vol.~\bibinfo{volume}{40}.
  \bibinfo{publisher}{ACM Press}, \bibinfo{address}{New York, New York, USA},
  \bibinfo{pages}{378--391}.
\newblock
\showISBNx{158113830X}
\showISSN{0362-1340}
\showDOI{%
\url{http://dx.doi.org/10.1145/1040305.1040336}}


\bibitem[\protect\citeauthoryear{Pelley, Chen, and Wenisch}{Pelley
  et~al\mbox{.}}{2014}]%
        {Pelley2014}
\bibfield{author}{\bibinfo{person}{Steven Pelley}, \bibinfo{person}{Peter~M.
  Chen}, {and} \bibinfo{person}{Thomas~F. Wenisch}.}
  \bibinfo{year}{2014}\natexlab{}.
\newblock \showarticletitle{{Memory persistency}}. In \bibinfo{booktitle}{{\em
  2014 ACM/IEEE 41st Int. Symp. Comput. Archit.}} \bibinfo{publisher}{IEEE},
  \bibinfo{pages}{265--276}.
\newblock
\showISBNx{978-1-4799-4394-4}
\showDOI{%
\url{http://dx.doi.org/10.1109/ISCA.2014.6853222}}


\bibitem[\protect\citeauthoryear{Pelley, Wenisch, Gold, and Bridge}{Pelley
  et~al\mbox{.}}{2013}]%
        {Pelley2013}
\bibfield{author}{\bibinfo{person}{Steven Pelley}, \bibinfo{person}{Thomas~F.
  Wenisch}, \bibinfo{person}{Brian~T. Gold}, {and} \bibinfo{person}{Bill
  Bridge}.} \bibinfo{year}{2013}\natexlab{}.
\newblock \showarticletitle{{Storage management in the NVRAM era}}.
\newblock \bibinfo{journal}{{\em Proc. VLDB Endow.\/}} \bibinfo{volume}{7},
  \bibinfo{number}{2} (\bibinfo{date}{oct} \bibinfo{year}{2013}),
  \bibinfo{pages}{121--132}.
\newblock
\showDOI{%
\url{http://dx.doi.org/10.14778/2732228.2732231}}


\bibitem[\protect\citeauthoryear{Prabhakaran, Bairavasundaram, Agrawal, Gunawi,
  Arpaci-Dusseau, Arpaci-Dusseau, Prabhakaran, Bairavasundaram, Agrawal,
  Gunawi, Arpaci-Dusseau, and Arpaci-Dusseau}{Prabhakaran
  et~al\mbox{.}}{2005}]%
        {Prabhakaran2005}
\bibfield{author}{\bibinfo{person}{Vijayan Prabhakaran},
  \bibinfo{person}{Lakshmi~N. Bairavasundaram}, \bibinfo{person}{Nitin
  Agrawal}, \bibinfo{person}{Haryadi~S. Gunawi}, \bibinfo{person}{Andrea~C.
  Arpaci-Dusseau}, \bibinfo{person}{Remzi~H. Arpaci-Dusseau},
  \bibinfo{person}{Vijayan Prabhakaran}, \bibinfo{person}{Lakshmi~N.
  Bairavasundaram}, \bibinfo{person}{Nitin Agrawal},
  \bibinfo{person}{Haryadi~S. Gunawi}, \bibinfo{person}{Andrea~C.
  Arpaci-Dusseau}, {and} \bibinfo{person}{Remzi~H. Arpaci-Dusseau}.}
  \bibinfo{year}{2005}\natexlab{}.
\newblock \showarticletitle{{IRON file systems}}. In \bibinfo{booktitle}{{\em
  Proc. Twent. ACM Symp. Oper. Syst. Princ. - SOSP '05}},
  Vol.~\bibinfo{volume}{39}. \bibinfo{publisher}{ACM Press},
  \bibinfo{address}{New York, New York, USA}, \bibinfo{pages}{206}.
\newblock
\showISBNx{1595930795}
\showISSN{0163-5980}
\showDOI{%
\url{http://dx.doi.org/10.1145/1095810.1095830}}


\bibitem[\protect\citeauthoryear{Qureshi, Srinivasan, Rivers, Qureshi,
  Srinivasan, and Rivers}{Qureshi et~al\mbox{.}}{2009}]%
        {Qureshi2009}
\bibfield{author}{\bibinfo{person}{Moinuddin~K. Qureshi},
  \bibinfo{person}{Vijayalakshmi Srinivasan}, \bibinfo{person}{Jude~A. Rivers},
  \bibinfo{person}{Moinuddin~K. Qureshi}, \bibinfo{person}{Vijayalakshmi
  Srinivasan}, {and} \bibinfo{person}{Jude~A. Rivers}.}
  \bibinfo{year}{2009}\natexlab{}.
\newblock \showarticletitle{{Scalable high performance main memory system using
  phase-change memory technology}}.
\newblock \bibinfo{journal}{{\em ACM SIGARCH Comput. Archit. News\/}}
  \bibinfo{volume}{37}, \bibinfo{number}{3} (\bibinfo{date}{jun}
  \bibinfo{year}{2009}), \bibinfo{pages}{24}.
\newblock
\showISBNx{978-1-60558-526-0}
\showDOI{%
\url{http://dx.doi.org/10.1145/1555815.1555760}}


\bibitem[\protect\citeauthoryear{Schwalb, Dreseler, Uflacker, and
  Plattner}{Schwalb et~al\mbox{.}}{2015}]%
        {Schwalb}
\bibfield{author}{\bibinfo{person}{David Schwalb}, \bibinfo{person}{Markus
  Dreseler}, \bibinfo{person}{Matthias Uflacker}, {and} \bibinfo{person}{Hasso
  Plattner}.} \bibinfo{year}{2015}\natexlab{}.
\newblock \showarticletitle{{NVC-Hashmap: A Persistent and Concurrent Hashmap
  For Non-Volatile Memories}}. In \bibinfo{booktitle}{{\em Proc. 3rd VLDB Work.
  In-Memory Data Mangement Anal. - IMDM '15}}. \bibinfo{publisher}{ACM Press},
  \bibinfo{address}{New York, New York, USA}, \bibinfo{pages}{1--8}.
\newblock
\showISBNx{9781450337137}
\showDOI{%
\url{http://dx.doi.org/10.1145/2803140.2803144}}


\bibitem[\protect\citeauthoryear{SNIA}{SNIA}{2013}]%
        {SNIA2013}
\bibfield{author}{\bibinfo{person}{SNIA}.} \bibinfo{year}{2013}\natexlab{}.
\newblock \bibinfo{title}{{NVM Programming Model (NPM)}}.
\newblock   (\bibinfo{year}{2013}).
\newblock
\showURL{%
\url{http://www.snia.org/sites/default/files/NVMProgrammingModel}}


\bibitem[\protect\citeauthoryear{Venkataraman, Tolia, Ranganathan, and
  Campbell}{Venkataraman et~al\mbox{.}}{2011}]%
        {Venkataraman2011}
\bibfield{author}{\bibinfo{person}{Shivaram Venkataraman},
  \bibinfo{person}{Niraj Tolia}, \bibinfo{person}{Parthasarathy Ranganathan},
  {and} \bibinfo{person}{Roy~H. Campbell}.} \bibinfo{year}{2011}\natexlab{}.
\newblock \showarticletitle{{Consistent and Durable Data Structures for
  Non-Volatile Byte-Addressable Memory}}.
\newblock \bibinfo{journal}{{\em Proc. 9th USENIX Conf. File Storage Technol. -
  FAST\/}} (\bibinfo{year}{2011}), \bibinfo{pages}{61--75}.
\newblock
\showISBNx{978-1-931971-82-9}


\bibitem[\protect\citeauthoryear{Volos, Tack, and Swift}{Volos
  et~al\mbox{.}}{2011}]%
        {Volos2011}
\bibfield{author}{\bibinfo{person}{Haris Volos}, \bibinfo{person}{Andres~Jaan
  Tack}, {and} \bibinfo{person}{Michael~M Swift}.}
  \bibinfo{year}{2011}\natexlab{}.
\newblock \showarticletitle{{Mnemosyne : Lightweight Persistent Memory}}.
\newblock \bibinfo{journal}{{\em Asplos\/}} (\bibinfo{year}{2011}),
  \bibinfo{pages}{1--13}.
\newblock
\showISBNx{9781450302661}
\showISSN{0362-1340}
\showDOI{%
\url{http://dx.doi.org/10.1145/1950365.1950379}}


\bibitem[\protect\citeauthoryear{Wang and Johnson}{Wang and Johnson}{2014}]%
        {Wang2014}
\bibfield{author}{\bibinfo{person}{Tianzheng Wang} {and} \bibinfo{person}{Ryan
  Johnson}.} \bibinfo{year}{2014}\natexlab{}.
\newblock \showarticletitle{{Scalable Logging through Emerging Non-Volatile
  Memory}}.
\newblock \bibinfo{journal}{{\em VLDB\/}} \bibinfo{volume}{7},
  \bibinfo{number}{10} (\bibinfo{year}{2014}), \bibinfo{pages}{865--876}.
\newblock


\bibitem[\protect\citeauthoryear{Wong, Lee, Yu, Chen, Wu, Chen, Lee, Chen, and
  Tsai}{Wong et~al\mbox{.}}{2012}]%
        {Wong2012}
\bibfield{author}{\bibinfo{person}{H.-S.~Philip Wong},
  \bibinfo{person}{Heng-Yuan Lee}, \bibinfo{person}{Shimeng Yu},
  \bibinfo{person}{Yu-Sheng Chen}, \bibinfo{person}{Yi Wu},
  \bibinfo{person}{Pang-Shiu Chen}, \bibinfo{person}{Byoungil Lee},
  \bibinfo{person}{Frederick~T. Chen}, {and} \bibinfo{person}{Ming-Jinn Tsai}.}
  \bibinfo{year}{2012}\natexlab{}.
\newblock \showarticletitle{{Metal-Oxide RRAM}}.
\newblock \bibinfo{journal}{{\em Proc. IEEE\/}} \bibinfo{volume}{100},
  \bibinfo{number}{6} (\bibinfo{date}{jun} \bibinfo{year}{2012}),
  \bibinfo{pages}{1951--1970}.
\newblock
\showDOI{%
\url{http://dx.doi.org/10.1109/JPROC.2012.2190369}}


\end{thebibliography}

\end{document}